\documentclass[conference]{IEEEtran}

\usepackage[T1]{fontenc} % Use 8-bit encoding that has 256 glyphs
\usepackage[utf8]{inputenc}
\usepackage[english]{babel} % English language/hyphenation
\usepackage{amsmath,amsfonts,amsthm, amssymb} % Math packages
\usepackage{mathtools}
\usepackage{acronym}
\usepackage{longtable}
\usepackage{textcomp}
\usepackage{graphicx}
\usepackage{acronym}
\usepackage{color}
\usepackage[linesnumbered]{algorithm2e}
\usepackage{epsfig}
\usepackage{url}
\usepackage{todonotes}
\usepackage{balance}
\usepackage[export]{adjustbox}
\usepackage{comment}
\usepackage{subfig}
\usepackage{pifont}
\usepackage{array}
\usepackage{pdfpages}
\usepackage{soul}

\newcommand{\proto}{PAST-AI}

\newcommand{\TODO}{\textcolor{blue}{\textbf{TODO}}}
\newcolumntype{P}[1]{>{\centering\arraybackslash}p{#1}}
\newcolumntype{M}[1]{>{\centering\arraybackslash}m{#1}}

\acrodef{LEO}{Low-Earth Orbit}
\acrodef{GPS}{Global Positioning System}
\acrodef{IMO}{International Maritime Organization}
\acrodef{GMDSS}{Global Maritime Distress and Safety System}
\acrodef{IoT}{Internet of Things}
\acrodef{IRA}{IRIDIUM Ring Alert}
\acrodef{PSK}{Phase Shift Keying}
\acrodef{BPSK}{Binary-Phase Shift Keying}
\acrodef{DQPSK}{Differentially-encoded Quadrature-Phase Shift Keying}
\acrodef{TMSI}{Temporary Mobile Subscriber Identity}
\acrodef{AI}{Artificial Intelligence}
\acrodef{ML}{Machine Learning}
\acrodef{CNN}{Convolutional Neural Network}
\acrodef{DNN}{Deep Neural Network}
\acrodef{RNN}{Recurrent Neural Network}
\acrodef{SVM}{Support Vector Machines}
\acrodef{SDR}{Software-Defined Radio}
\acrodef{COTS}{Commercial Off-The-Shelf}
\acrodef{ADS-B}{Automatic Dependent Surveillance - Broadcast}
\acrodef{SNR}{Signal-to-Noise Ratio}
\acrodef{DCTF}{Differential Constellation Trace Figure}
\acrodef{RF}{Radio-Frequency}
\acrodef{DCNN}{Deep Convolutional Neural Network}
\acrodef{AUC}{Area Under the Curve}
\acrodef{PLL}{Phased Locked Loop}
\acrodef{ROC}{Receiver Operating Characteristic}
\acrodef{m.s.e.}{mean square error}
\acrodef{FP}{False Positive}
\acrodef{FN}{False Negative}
\acrodef{TPR}{True Positive Rate}
\acrodef{FPR}{False Positive Rate}

\begin{document}

\title{PAST-AI: Physical-layer Authentication of \\Satellite Transmitters via Deep Learning}

\author{
    \IEEEauthorblockN{Gabriele Oligeri, Simone Raponi, Savio Sciancalepore, Roberto Di Pietro} \\
    \IEEEauthorblockA{Division of Information and Computing Technology \protect\\ College of Science and Engineering, Hamad Bin Khalifa University - Doha, Qatar
    \\ \{goligeri, sraponi, ssciancalepore, rdipietro\}@hbku.edu.qa }
}

%\author{Anonymous Author(s)}

\maketitle

% Page numbering
\thispagestyle{plain}
\pagestyle{plain}

\begin{abstract}
    Physical-layer security is regaining traction in the research community, due to the performance boost introduced by deep learning classification algorithms. 
    This is particularly true  for sender authentication in wireless communications via radio fingerprinting.  
    %Indeed, Very recent results showed that deep learning classifiers, such as \acf{CNN} and autoencoders, can be successfully adopted to mitigate the impact of noise, this latter one being the  main barrier %which usually prevents     for the  receivers to detect and extract the features (fingerprint) enabling the authentication of the hardware at the transmitting source.
    However, previous research efforts mainly focused on terrestrial wireless devices while, to the best of our knowledge, none of the previous work took into consideration satellite transmitters. 
    The satellite scenario is generally challenging because, among others, satellite radio transducers feature non-standard electronics (usually aged and specifically designed for harsh conditions). Moreover, the fingerprinting task is specifically difficult for Low-Earth Orbit (LEO) satellites (like the ones we focus in this paper) since they orbit at about $800$~Km from the Earth, at a speed of around $25,000$~Km/h, thus making the receiver experiencing a down-link with unique attenuation and fading characteristics.
    In this paper, we propose \proto, a methodology tailored to authenticate LEO satellites through fingerprinting of their IQ samples, using advanced AI solutions. Our methodology is tested on real data---more than $100M$~I/Q samples--- collected from an extensive measurements campaign on the IRIDIUM LEO satellites constellation, lasting $589$ hours. %In particular, we focused , collecting  589 hours of measurements, totaling  more than 100M of IQ samples. 
    Results are striking: we prove that Convolutional Neural Networks (CNN) and autoencoders (if properly calibrated) can be successfully adopted to authenticate the satellite transducers, with an accuracy spanning between $0.8$ and $1$, depending on prior assumptions. 
    
    The proposed methodology, the achieved results, and the provided insights, other than being interesting on their own, when associated to the dataset that we made publicly available, will also pave the way for future research in the area.
\end{abstract}

% \keywords{
% IRIDIUM, Satellite Authentication, Spoofing Detection, IQ Fingerprinting, Convolutional Neural Networks, Autoencoders.
% }

%\begin{keywords}
%\end{keywords}

\section{Introduction}
\label{sec:introduction}

Physical-layer authentication relies on detecting and identifying unique characteristics embedded in over-the-air radio signals, thus enabling the identification of the hardware of the transmitting source~\cite{wang2016_commag,xiao2007_icc}. 
Wireless Physical-layer authentication is also known as radio fingerprinting when referring to the challenge of both detecting and extracting features from the received signal (fingerprint), which can uniquely identify the transmitting source~\cite{xu2016_comst, ibrahim2020_tecs}. 

Physical-layer authentication can significantly enhance the security and privacy of wireless channels in two adversarial scenarios: (i) spoofing; and, (ii) replay attacks. 
The former involves a rogue transmitting source attempting to impersonate a legitimate one, while the latter assumes the adversary being able to re-transmit previously eavesdropped messages~\cite{schmidt2016_csur}. Despite spoofing detection can be achieved by authenticating the transmitting source with standard cryptographic techniques (e.g., digital signatures), in many scenarios involving  massive deployments (e.g., IoT),  difficult to reach devices (e.g., satellites), or when the cryptography-induced overhead is considered excessive,  digital signatures might be inefficient~\cite{soltanieh2020_jrfi}. Alternative solutions could involve crowd-sourcing, i.e., cross-checking context information to validate the transmitting source~\cite{oligeri_wisec_2019,oligeri_wisec_2020}. 
Replay attacks can be even more difficult to detect, being dependent on specific protocol flaws: the adversary re-transmits encrypted information, which will be considered as valid if not timestamped. Both spoofing and replay attacks can be prevented if the receiver can authenticate the hardware of the transmitting source~\cite{zhou2019_cns}.

Many researchers have already undertaken the challenge of extracting %collision-free 
fingerprints and developing effective detection algorithms to extract and match the fingerprints (see Sec.~\ref{sec:related_work} for an overview). The cited tasks have  been mainly achieved by resorting to dedicated hardware at the receiver side, featuring high sampling resolution and better signal quality. Indeed, \acp{SDR} played a major role as an enabling technology for radio fingerprinting. Specifically, \acp{SDR} provide both high-resolution bandwidth (thus exposing the features of the transmitting source) and high signal-to-noise ratio (thus facilitating the extraction of the features to the back-end algorithms). Unfortunately, radio noise still represents the major issue for all the state-of-the-art solutions. Indeed, the fingerprint of the transmitting source is mixed---drown, in many cases---with the noise of the radio channel. Therefore, discriminating between the needed features and the noise brings back the problem of developing effective algorithms to achieve the cited objective.

Recently, \acp{CNN} have been adopted for radio fingerprinting in several scenarios, such as ADS-B, WiFi, and Zigbee, to name a few~\cite{yu2019_wimob,sankhe2020_tccn,ying2019_cns,shawabka2020_infocom}. The idea behind the adoption of \acp{CNN} relies on exploiting their multidimensional mapping during the learning process to detect and extract reliable radio fingerprints. However, all of the recent contributions took into account terrestrial links, only.

Although achieving interesting performance, there are still some open fundamental questions related to \acp{CNN}, such as the intrinsic time-stationarity nature of the \acp{CNN} and how the wireless channel (in terms of attenuation and fading) affects the learning and detection processes~\cite{shawabka2020_infocom}.
%, and finally, how fingerprints are affected by long-term deployments (with changing of environmental temperature and hardware aging).
%RDP no need for having it here, e teniamolo in dispensa
Recent results~\cite{shawabka2020_infocom} based on real measurements on terrestrial wireless links confirmed that the wireless channel significantly impacts the classification accuracy (up to $80\%$), thus confirming the need for more effective classification techniques. It is worth noting that no prior contribution has been made up to date to physical layer authentication of satellite transmitters (in particular the IRIDIUM constellation), given their intrinsic challenges. Indeed, LEO satellites, which IRIDIUM constellation is part of, are characterized by unique features: the satellite transmitter is at around 800Km from earth, and moves at about 7Km/s with a pass duration of about 8 minutes~\cite{oligeri_wisec_2020}---involving a radio link (quality) that significantly changes over the time. Indeed, we observe that attenuation and multi-path fading can significantly change when the satellite is either on top of the receiver or far away, just over the horizon (before disappearing). Therefore, the noise affecting the satellite link makes radio fingerprinting in satellite a unique, more challenging scenario,  requiring additional research.

{\bf Contribution.} This paper provides the following contributions: 
\begin{itemize}
    \item We push further the current state-of-the-art in physical-layer authentication, by proposing \proto, i.e., a set of new methodologies specifically designed to perform radio fingerprinting over LEO satellite links. 
    \item We propose a new technique to represent  IQ samples in  input to  AI classification algorithms.
    \item We prove that \acf{CNN} and autoencoders can be effectively adopted to fingerprint radio satellite transmitters. 
    \item We propose two different classification scenarios, i.e., \emph{intra-constellation satellite authentication} and \emph{satellite authentication in the wild}, which fit the adopted classification algorithm and their assumptions. 
    \item We provide several insights to properly calibrate the algorithm parameters, achieving overwhelming performance, i.e., an accuracy greater than 0.8 for the former scenario and average \ac{AUC} equal to 1 for the latter (vast majority of the satellites).
\end{itemize}
{\bf Paper organization.} The rest of this paper is organized as follows. Section~\ref{sec:related_work} reviews related work on physical-layer fingerprinting; Section~\ref{sec:background} introduces background details on IQ modulation, AI techniques, and the IRIDIUM satellite constellation;
Section~\ref{sec:iq_processing} illustrates the data acquisition campaign and the initial data processing; Section~\ref{sec:satellite_authentication} introduces the \proto\ methodology; Section~\ref{sec:intra-satellites} focuses on the intra-constellation satellite authentication scenario; Section~\ref{sec:satellite_authentication_inthe_wild} details the authentication scenario with minimal satellites' knowledge; and, finally, Section~\ref{sec:conclusion} tightens the conclusions.

\section{Related Work}
\label{sec:related_work}

Physical-layer authentication solutions based on the analysis of raw IQ samples have gained significant popularity in the last years, and have been adopted in a variety of scenarios and communication technologies.

For instance, in the context of mobile cellular networks, the authors in~\cite{zhuang2018_asiaccs} proposed \emph{FBSLeuth}, a framework able to identify rogue 2G mobile cellular base stations by analyzing the hardware impairments of the transmitting devices, such as the error vector magnitude of the signals, the phase error, the frequency error, the IQ offset, the IQ quadrature skew, and the IQ imbalance. To identify the rogue base stations, they used supervised \ac{ML} techniques, specifically the \ac{SVM} classification algorithm. In the same context, the authors in~\cite{wang2020_infocom} relied on \ac{DCTF}-based features and \acp{CNN} to identify mobile phones. Specifically, the authors used image discrimination techniques to discriminate among six ($6$) different mobile phones, with outstanding accuracy and a reduced observation window.

In the context of WiFi, the authors in~\cite{sankhe2019_infocom} first were able to distinguish among \ac{COTS} WiFi devices and \acp{SDR} emitting similar WiFi-compliant signals. Specifically, using a \ac{CNN}-based architecture operating on raw IQ samples, they could identify precisely among sixteen ($16$) \acp{SDR}s. The authors further extended their work in~\cite{sankhe2020_tccn}, showing how the classification accuracy can reach over $99$\% by smartly removing the noise effects of the wireless channel.

The impact of the wireless channel on wireless radio fingerprinting has been specifically studied by the authors in~\cite{shawabka2020_infocom}. They evaluated the accuracy of \ac{CNN}-based methods in several operating conditions, i.e., in an anechoic chamber, in the wild, and using cable connections, investigating both WiFi and \ac{ADS-B} signals (employed in the aviation domain). They revealed that the wireless channel can severely affect the accuracy of the radio fingerprinting, degrading the classification accuracy up to the $85$\% in low-\ac{SNR} regime. At the same time, they showed that equalizing IQ data can slightly enhance the quality of the fingerprinting, when possible. Similar results and findings were achieved also by the authors in~\cite{jian2020_iotmag}. By working on the same dataset, the authors confirmed that partial equalization of the samples can improve the accuracy of the \ac{CNN}-based architecture in identifying the transmitter, while the accuracy generally decreases with the decrease of the \ac{SNR}.

\ac{ADS-B} signals have been investigated also by the authors in~\cite{ying2019_cns}, by using an autonomously-made dataset. Specifically, the authors compared the performance of three different \acp{DNN}s, characterized by a different number of hidden layers and nodes (i.e., neurons), and they showed that the performance of the classifiers slightly decreases when the number of considered aircraft increases, as well as by reducing the training set ratio.

The IQ fingerprinting technique is particularly promising for the \ac{IoT} domain, as it could avoid the installation of dedicated cryptography techniques on memory-limited and computationally-constrained devices. These considerations motivated several studies, applying IQ fingerprinting techniques on \ac{IoT} devices. For instance, the authors in~\cite{jafari2018_milcom} relied on multiple deep learning models, i.e., \ac{CNN}, \ac{DNN}, and \ac{RNN}, to discriminate among six ($6$) identical Zigbee devices, showing that the \ac{DNN} model slightly outperforms the others, especially with short windows sizes. The same number of devices has been adopted also by the authors in~\cite{bassey2019_fmec}, which used \acp{CNN}, dimensionality reduction, and de-correlation to further improve the performance of the classification task for \ac{IoT} devices. 

Recently, the authors in~\cite{yu2019_wimob} demonstrated that stacked autoencoders can be used to enhance the performance of \ac{CNN}-based methods for IQ fingerprinting, especially in low-\ac{SNR} scenarios. To verify their findings, they used twenty-seven ($27$) CC2530 micro-controllers, and they were able to distinguish each of them with accuracy over $90$~\% starting from $5$~dB \ac{SNR}.

Another recent contribution is provided in~\cite{balakrishnan2020_tifs}, where the authors identified mm-WAVE transmitters operating at the frequency of $60$~GHz by analyzing the spatio-temporal information of the beam patterns created by the antennas in the array. 

Despite the significant number of contributions in the field of IQ fingerprinting, the satellite scenario has not yet been considered, thus still representing a challenging research problem. Indeed, being the satellite located at a significant altitude from the ground, the signals are typically characterized by a low \ac{SNR} and a significant noise level, thus making the fingerprinting task more challenging. 

At the time of this writing, the only contribution working on the fingerprinting of satellites is~\cite{foruhandeh2020_wisec}. The authors argue to be able to identify \ac{GPS} spoofing attacks by analyzing the received IQ samples, by using a statistical approach based on scores computed over characterizing Multi-Variate Normal (MVN) distributions. However, they extracted the IQ samples after the IQ demodulation at the \ac{RF} front-end, and specifically after the E-P-L correlators in the receiving chain. Therefore, their solution does not act on raw IQ samples, and applies only to US \ac{GPS} satellites. Finally, note that the authors focused on the detection of GPS spoofing attacks, and they distinguish \acp{SDR} from legitimate satellites, not the specific transmitting satellite. Conversely, in this paper we identify the specific satellite transmitting an IRIDIUM signal, considering \emph{raw} IQ samples, before any demodulation operation. As a result, our methodology applies to a wider set of scenarios than spoofing attacks, and it is potentially applicable to all \ac{LEO} satellites constellations adopting \ac{PSK} modulation techniques.

\section{Background}
\label{sec:background}
In this section we revise the technical background providing the needed information that will be leveraged in next sections.

\subsection{IQ (de)modulation}
\label{sec:background_iq}

\begin{figure}[htbp]
    \centering
    \includegraphics[width=\columnwidth]{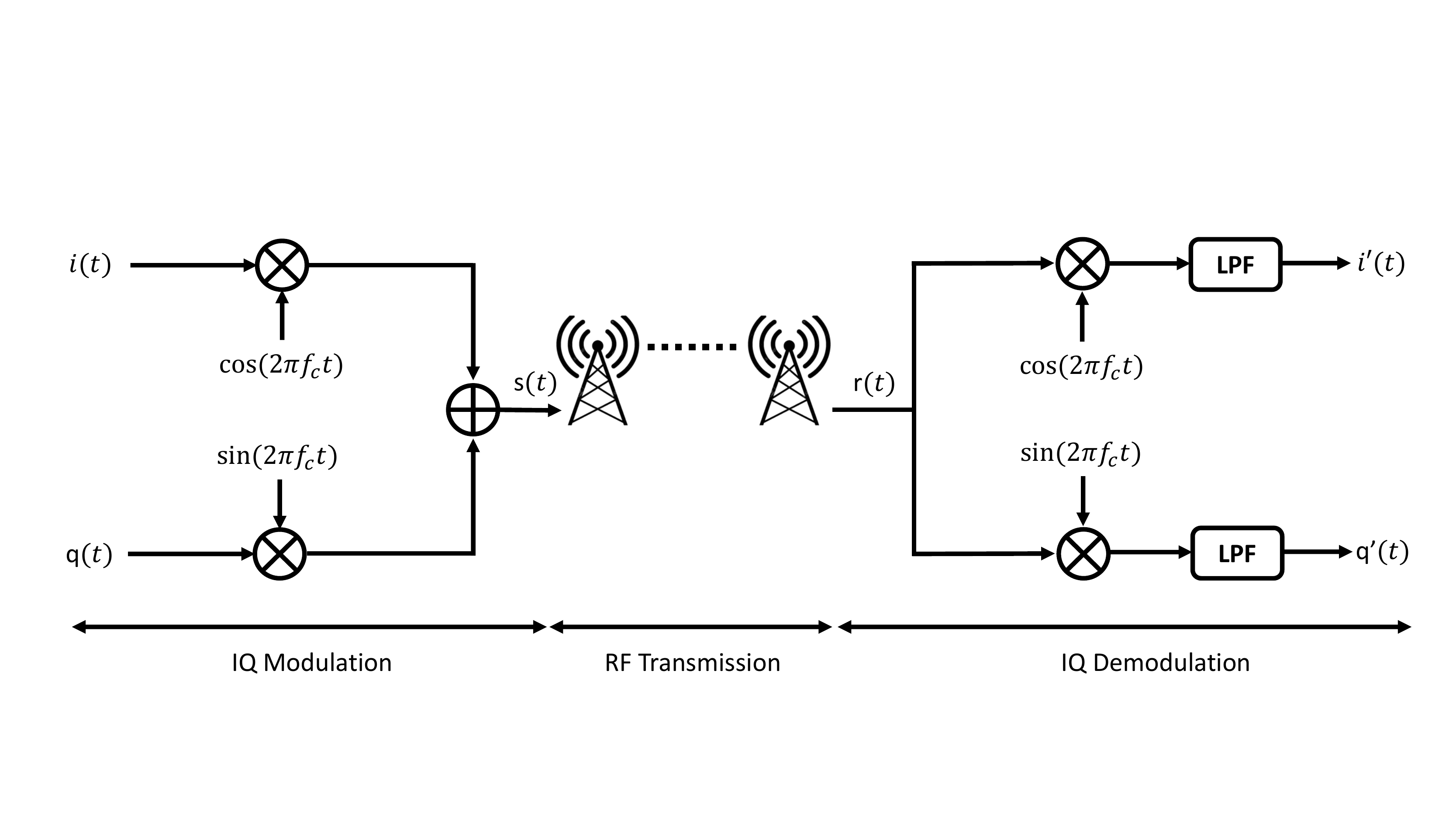}
    \caption{Modulation and Demodulation of a digital signal represented by its phase $i(t)$ and quadrature $q(t)$ components.}
    \label{fig:txrxiq}
\end{figure}

Digital modulation schemes involve the processing of a (low frequency) baseband signal, i.e., a bit sequence $b_i \in \{0, 1\}$ with $i \in [1, N]$, to make it suitable for the transmission virtually anywhere in the RF spectrum (high frequency).
Several techniques have been developed to achieve the aforementioned result, but \emph{IQ modulation} is the most adopted due to practicality: 
%hardware concerns, i.e., 
efficient IQ (de)modulators are available as inexpensive System on Chip (SoC) technology. Figure~\ref{fig:txrxiq} shows the block diagram of a typical communication system involving IQ modulation, RF transmission, and IQ demodulation. 
According to the scheme, a sequence of bits should be preliminary converted into \emph{IQ symbols}, i.e., $i(t)$ and $q(t)$ in Fig.~\ref{fig:txrxiq}. Different families of modulation schemes are possible, e.g., Amplitude Shift Keying (ASK), Frequency Shift Keying (FSK), or Phase Shift Keying (PSK), depending on how the sequence of bits is converted to the {\em in-phase} $i(t)$ and {\em quadrature} $q(t)$ components (recall Fig.~\ref{fig:txrxiq}). 

\begin{figure}[htbp]
    \centering
    \includegraphics[width=\columnwidth]{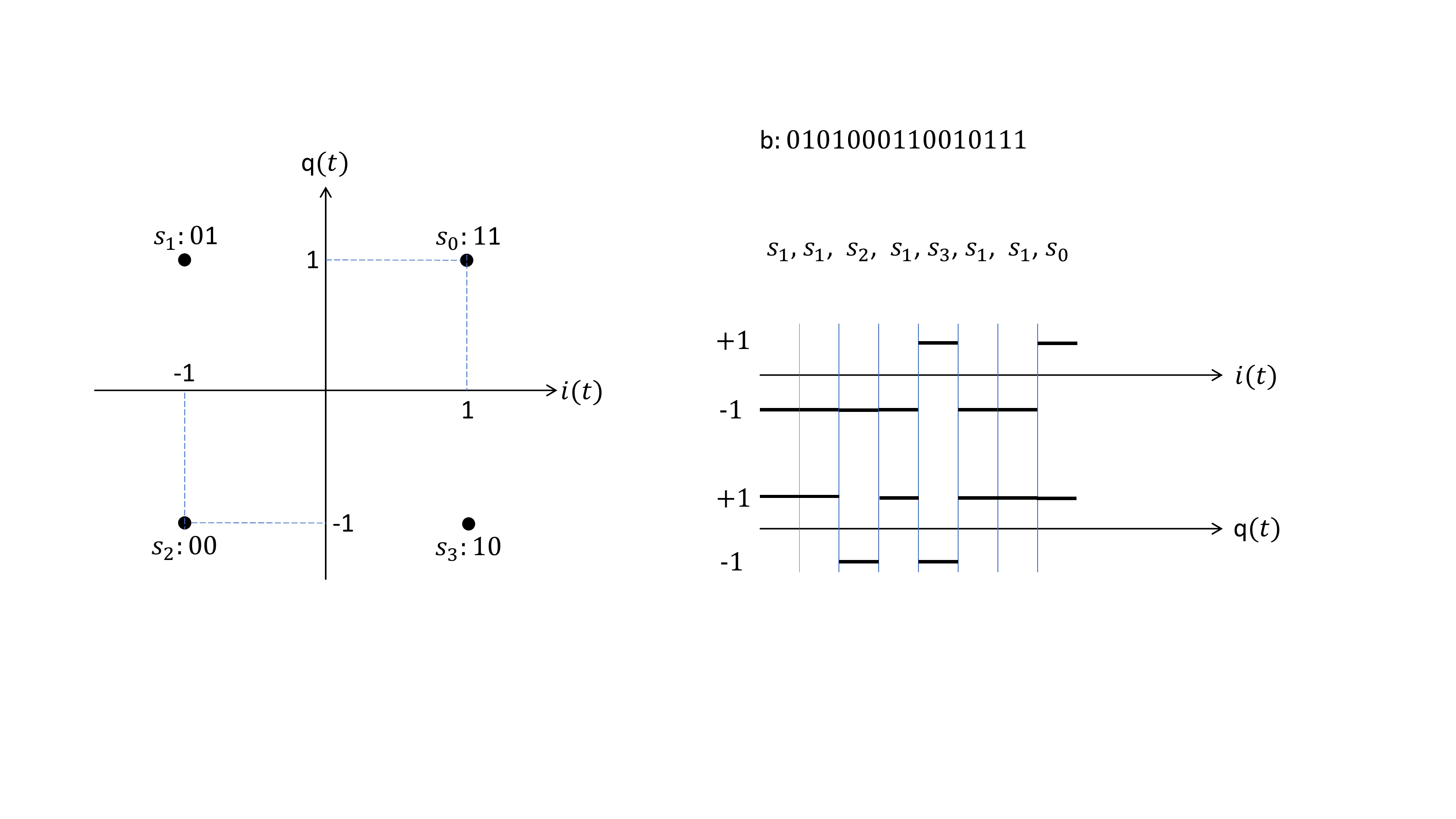}
    \caption{Quadrature Phase Shift Keying (QPSK) mudulation example: from bit sequence $b$ to the in-phase $i(t)$ and quadrature $q(t)$ components.}
    \label{fig:bits_to_symbols}
\end{figure}

As a toy example, we consider the Quadrature Phase Shift Keying (QPSK or 4-PSK)---the one adopted by Iridium is very similar and we will discuss it in the next sections. 
QPSK maps pair of bits into (four) IQ symbols, i.e, $\{1, 1\} \rightarrow s_0$, $\{0, 1\}\rightarrow s_1$, $\{0, 0\}\rightarrow s_2$, and $\{1, 0\}\rightarrow s_3$, as depicted by Fig.~\ref{fig:bits_to_symbols}. It is worth noting that the aforementioned mapping can be easily achieved by setting $i(t) = \{-1, 1\}$ and $q(t) = \{-1, 1\}$, as depicted in Fig.~\ref{fig:bits_to_symbols}. For instance, the bit string $b:[0, 1, 0, 1, 0, 0, 0, 1, 1, 0, 0, 1, 0, 1, 1, 1]$ becomes the sequence of symbols $[s_1, s_1, s_2, s_1, s_3, s_1, s_1, s_0]$, thus obtaining the in-phase $i(t)$ and quadrature $q(t)$ signal components. For the sake of completeness, we highlight that both $i(t)$ and $q(t)$ should be subject to other filtering stages and they cannot be directly used as mentioned in Fig.~\ref{fig:txrxiq}, since the sharp level changes will eventually cause $s(t)$ to have a very large bandwidth \cite{rappaport}.

Let us now complete the discussion about the IQ (de)modulation previously introduced by Fig.~\ref{fig:txrxiq}. $i(t)$ and $q(t)$ components are \emph{modulated} adopting an in-phase ($\cos{2\pi f_c t}$) and a quadrature ($\sin{2\pi f_c t}$) signal at the reference frequency $f_c$ (\emph{carrier}). The resulting signals are summed up to obtain $s(t)$, the actual RF signal. Figure~\ref{fig:txrxiq} takes into account any propagation phenomena, such as fading and attenuation, that may affect the received signal, and therefore $r(t) \neq s(t)$.
%takes into account that the received signal will be affected by propagation phenomena such as fading and attenuation, and therefore, $r(t) \neq s(t)$. 
The demodulation block is the reciprocal of the modulator. Indeed, the received signal $r(t)$ is multiplied by both an in-phase and a quadrature signal at frequency $f_c$, and then, low pass filtered in order to remove the unwanted upper sidebands. The final result consists of $i'(t)$ and $q'(t)$ that can be arbitrarily different from the original $i(t)$ and $q(t)$ signal components. The greatest source of difference usually comes from RF propagation, which can affect $i(t)$ and $q(t)$ so badly to make the symbol recovery impossible. When the \emph{signal-to-noise ratio} is large enough, the symbols are evenly distributed and the information recovery becomes feasible. 

Further, there are also minor effects that introduce small offset in the IQ symbols. A typical example is constituted by impairments and biases introduced by small differences in the electronics components, that, although being mass produced by controlled and standardized assembly lines, are still characterized by imperfections at nano-scale, that affect 
%become ``visible" in 
the displacement of the symbols. The analysis introduced in latter sections proves that the symbols' displacement is systematic, thus being at least theoretically possible to detect it, measure it, and eventually leverage it to identify the hardware causing it.
We moved from theory to practice, showing a viable method leveraging AI to achieve the cited objective.

\subsection{Deep Learning classifiers and Transfer Learning}
\label{sec:background_ai}

In this subsection, the Deep Learning classifiers adopted in this study are introduced, together with the transfer learning technique, that allowed us to notably improve the accuracy during the multi-class classification task.  

\subsubsection{Autoencoders}
\label{sec:autoencoders}

%An autoencoder is defined as an artificial neural network trained to find a (usually compressed) representation (i.e., encoding) of a training set with the aim of accurately reconstructing the input data by solely relying on such encoding. 

An autoencoder is defined as an artificial neural network whose goal is to learn an optimal representation (i.e., encoding) of a training set from which it is possible to accurately reconstruct the input data. 
Although it may seem trivial (i.e., the mere copy of the input data to the output may easily lead to an outstanding accuracy), to identify useful features, the internal function responsible for the research of good encoding candidates is usually constrained. For instance, the autoencoder may be forced to find an encoding smaller than the input data (i.e., undercomplete autoencoder). 

Traditionally, this unsupervised technique has been widely adopted to perform dimensionality reduction and feature learning, since it may be tuned to generate smaller encodings as similar as possible to the original input, while recently autoencoders are also being put to the forefront of generative modeling~\cite{goodfellow_deep_learning}.

The more similar the output reconstructed starting from such encoding is to the training set, the more likely the autoencoder is said to be able to represent input data. In case the encoding is (parametrically) smaller than the input data, the feature reduction phase is successful.

The basic autoencoder model has been quickly followed by many variants, each one forcing the learned encoding to boast a different property. Valuable examples are the regularized autoencoders, able to learn the most salient features of the data distribution~\cite{goodfellow_deep_learning}, and variational autoencoders, able to provide a framework to learn deep latent-variable models as well as the corresponding inference models~\cite{kingma2019introduction}. 

An autoencoder usually consists of four main components: (i) an encoder, that allows the model to learn how to represent the features of the input data; (ii) a bottleneck, identified as the layer containing the encoding of the training set; (iii) a decoder, that allows the model to learn how to reconstruct the input data from the encoding; and, (iv) the reconstruction error function, useful to measure the performance of the model during the whole training. 

The performance offered by the autoencoders positively impacted their wide applications, which now range from intrusion detection tasks~\cite{ieracitano2020neurocomputing}, to anomaly detection~\cite{nazir2020autoencoder}, and DDoS attack detection~\cite{yang2020ieeeifip}.
%which now range from face recognition tasks~\cite{gao2015single} to the vector representations finding for audio segments~\cite{chung2016audio}. 

In this paper, we rely on autoencoders to perform the one-class classification task on the IRIDIUM satellites. 
The intuition behind the adoption of autoencoders to face such a challenge is the following: starting from a distribution (i.e., class) $X$, the reconstruction of input data drawn from the same distribution $X$ is easier (i.e., the error metric is reduced) than the reconstruction of input data drawn from any other distribution $Y$, with $Y \ne X$.

\subsubsection{Convolutional Neural Networks}
\label{sec:cnn}
A \ac{CNN} is defined as a \ac{DNN} that boasts at least one convolutional layer, i.e., a layer performing convolutional operations. A convolutional operation, in turn, is the mathematical combination of two functions that produces a third function, being the expression of the change of shape caused by the application of one function to the other. 
In the case of \ac{CNN}, a convolution consists of a slide of a parametric-sized filter (also known as operator) over the input representation. Being the filter smaller compared to the input representation, it is applied to different overlapping portions of the input, thus generating a feature map. Different filters allow to catch different patterns within the input representation (i.e., in case the input is represented as an image, operators can be used to highlight edges, corners, and possibly other patterns).

A typical \ac{CNN} is composed of three types of layers: (i) convolutional layers, to build the feature map of the input representation; (ii) pooling layers, to reduce the number of learnable parameters and discretize the input; and, (iii) fully connected layers, usually representing the last layers of the architecture, to hold the high-level features found during the convolutions and to learn non-linear combinations of them.

When compared to multi-layer perceptrons, \acp{CNN} present characteristics that discourage the learning of too complex and expensive models, thus being recognized as their regularized version (i.e., a version that allows containing overfitting by construction). 
Indeed, while in multi-layer perceptrons several fully connected layers (i.e., layers whose neurons are fully connected to the ones of the next layer) are employed to perform classification, \ac{CNN}s exploit a hierarchical structure able to learn complex patterns by relying on the combination of small and simple ones~\cite{raponi2020sound}. The reduced number of connections and parameters made \ac{CNN}s extremely appreciable in several domains due to their ability to be trained quickly and more accurately than previous feed-forward models. 
Specifically, applications can be found in handwriting recognition, face detection, behavior recognition, recommendation systems, speech recognition, image classification, and Natural Language Processing~\cite{liu2017survey}.

\subsubsection{Transfer Learning}
\label{sec:transfer_learning}

Until a few years ago, conventional machine learning algorithms have been designed to work in isolation, trained from scratch every single time to solve specific tasks. However, training a network from scratch may be cumbersome, since the available datasets may not be rich enough to effectively capture the features. As a result, the resulting classifier could not generalize properly when applied in the wild.

With the introduction of transfer learning, however, the learning phase of the algorithms has been completely revolutionized. The general idea of transfer learning is to take advantage of the knowledge learned while solving a task of a particular domain to simplify the learning phase for a related domain task. In this paper, in order to perform multi-class classification on the IRIDIUM satellites, we exploited the knowledge of the Resnet-18 \ac{CNN}, pre-trained on the popular ImageNet dataset. Resnet, introduced by Microsoft researchers in 2015, proved to be the most performant \ac{CNN}, since it is structured in such a way to allow achieving deeper architectures with a reduced number of parameters~\cite{he2016_cvpr}. Details on the ResNet-18 \ac{CNN} and the transfer learning methodology adopted in this study (e.g., fine-tuning or freezing-layers) are detailed in Section~\ref{sec:convolutional_neural_networks}.

\subsection{Iridium Satellite Constellation}
\label{sec:iridium_constellation_and_data_acquisition}

The IRIDIUM satellite constellation was conceived in 1987, and first operated in 1993 by IRIDIUM SSC, founded by Motorola~\cite{pratt1999_comst}. The constellation is constituted by a set of \ac{LEO} satellites, orbiting $800$~km above the Earth surface, and arranged so that they can guarantee full Earth coverage at any time. The name of the satellite constellation is inspired by the originally-planned number of satellites, i.e., $77$, coincident with the atomic number of the IRIDIUM chemical element. However, to minimize deployment costs while still guaranteeing Earth coverage, only $66$ satellites are operational nowadays.

IRIDIUM radio signals are transmitted in the L-band, in the frequency range $[1,616 - 1,626.5]$~MHz. At the ground, IRIDIUM subscribers can receive such signals as well as  transmitting by using dedicated mobile satellite devices, provided by companies such as Motorola and Kyocera. Today, IRIDIUM is mainly used on-board of vessels, to initiate and receive calls when located off-shore. In this context, starting from January 2020, the \ac{IMO} has certified IRIDIUM as an approved \ac{GMDSS} service provider for vessels. However, IRIDIUM transceivers are also used in the aviation, railway, and critical infrastructures domain, and recently they have received significant attention also in the emerging satellite-\ac{IoT} application domain~\cite{iridium_iot}.

Each IRIDIUM satellite includes an array of antennas, hereby referred to as \emph{beams}, that widens the transmission range of the satellite at the ground. Overall, each satellite has $48$~beams and an additional antenna dedicated to the identification of the satellite. Note that the transmission power adopted by the \emph{satellite} antenna is higher than the one used by the \emph{beams}, so that any receiver that could decode the signal emitted by a beam can also receive the information about the satellite itself. %Many interesting information about the speed of the satellites, the coverage at the ground, the arrangement of the beams, and the satellite pass duration can be found in the reference paper~\cite{oligeri_wisec_2020}.

Overall, two channels categories are available, i.e., \emph{system overhead channels} and \emph{bearer service channels}. In this paper, we focus our attention on one of the \emph{system overhead channels}, i.e., the \ac{IRA} broadcast channel. It is a broadcast, unencrypted, downlink-only channel, operating at the center frequency $1,626.27$~MHz, and used to deliver information useful for handover operations at the ground. \ac{IRA} messages are characterized by a $12$~bytes preamble, encoded according to the \ac{BPSK} modulation scheme, while the rest of the information ($103$~bytes) follows the \ac{DQPSK} modulation. Such information include the ID of the satellite emitting the packet, the specific transmitting beam (the beam ID is $0$ in the case the transmitter is the one identifying the satellite), the position of the satellite (expressed in latitude, longitude, and altitude), and further information used for handover, e.g., the \ac{TMSI} of any user subject to handover. Note that \ac{IRA} packets can have different sizes, depending on the amount of \acp{TMSI} included in the message, as well as the presence of additional specific paging information. 

Previous contributions such as~\cite{oligeri_wisec_2020} used the information included into the IRA messages to reverse-engineer several system parameters of the IRIDIUM constellation, such as the speed of the satellites, the coverage at the ground, the arrangement of the beams, and the satellite pass duration. 
In this paper, we further extend those results, by providing additional hints on the time needed to \emph{observe} a specific satellite, the distribution of IQ samples, the effect of the noise, and the expected number of IQ samples per satellite pass (see Section~\ref{sec:iq_processing}. All these information are instrumental to the scope of our work, i.e., the authentication of the IRIDIUM satellite at the physical-layer, by using raw IQ samples.

\section{IRIDIUM Data Acquisition and Processing}
\label{sec:iq_processing}

In this section, we first describe the equipment (hardware and software) that has been adopted for our measurement campaign, later we depict how we reverse-engineered the architectural parameters of the IRIDIUM satellite constellation and, finally, we introduce how we exploited the IQ samples to authenticate the satellite transmitters.

\subsection{Measurement Set-up}
\label{sec:measurement_setup}

\begin{figure}[htbp]
    \centering
    \includegraphics[width=\columnwidth]{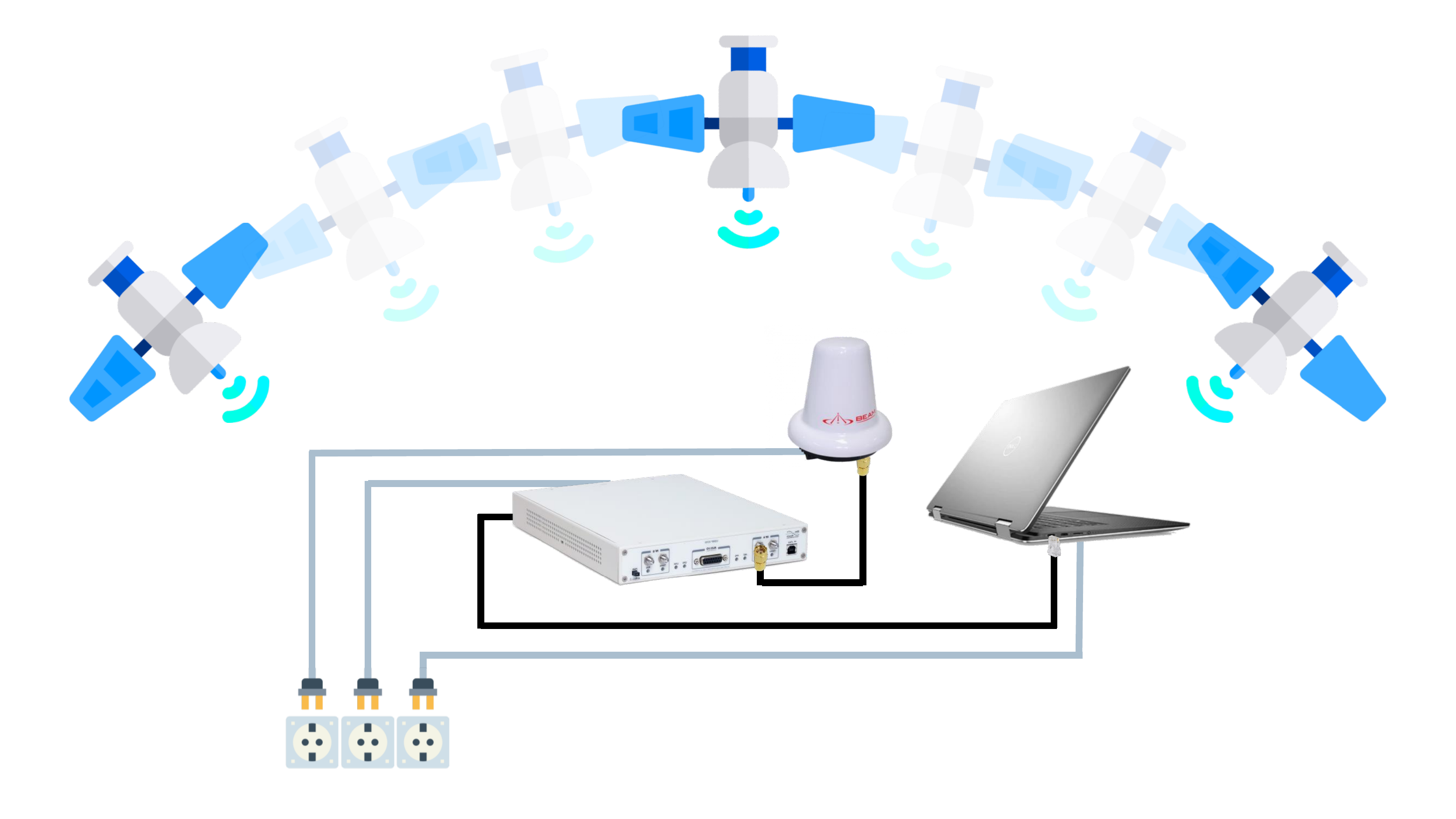}
    \caption{Measurement Setup: we adopted an active (pre-amplified) Iridium antenna (Beam RST740) connected to a USPR X310 Software Defined Radio.}
    %RDP3 nessun riferimento al laptop?
    \label{fig:scenario}
\end{figure}

The measurement setup is illustrated in 
Figure~\ref{fig:scenario}.

% IRIDIUM Antenna - Ettus X310 - Laptop - GNURadio
The hardware used to acquire IRIDIUM signals consists of a dedicated L-Band IRIDIUM antenna, connected to a general-purpose Ettus Research X310 \ac{SDR}. The antenna is an IRIDIUM Beam Active Antenna, model RST740, commonly used by commercial IRIDIUM transceiver~\cite{iridium_antenna}. The antenna is connected through an SMA cable to the Ettus X310 SDR~\cite{ettus}, integrating the UBX160 daughterboard~\cite{ubx}. In turn, the SDR is connected via Ethernet to a Laptop Dell XPS15 9560, equipped with 32GB of RAM and 8 Intel Core i7700HQ processors running at 2.80 GHz. 

% gr-iridium module, modification to take IRA messages after PLL
On the software side, we used the well-known GNURadio development toolkit. Specifically, we adopted the \emph{gr-iridium} module to detect and acquire IRIDIUM messages~\cite{iridiumgr}. In addition,
we used the \emph{iridium-toolkit} tool to parse \ac{IRA} messages~\cite{iridium-toolkit}. In detail, we modified the \emph{gr-iridium} module in a way to log the IQ samples of all the \emph{valid} IRIDIUM packets, i.e., the ones containing the 12~bytes BPSK-modulated preamble, typical of the IRIDIUM messages. For each of these packets, we logged the values of the IQ samples after the filtering and synchronization performed by the \ac{PLL}. Next, we used the \emph{iridium-toolkit} tool to log only valid \ac{IRA} packets. 

% data excerpt
Our measurement campaign has been carried out in very harsh conditions, i.e., by exposing the IRIDIUM antenna out of the window of an apartment. 
This is a worst-case scenario, since part of the open sky is obstructed by the wall of the building, attenuating and deviating the signal coming from the satellites. However, we highlight that this is not a limitation of our study. Conversely, the high-level performance achieved in such a disadvantaged scenario paves the way for further improvement. 

Overall, we continuously acquired IRIDIUM signals for about $589$~hours (24 days), gathering a total number of $102,318,546$ IQ samples ($1,550,281$ per satellite, on average). An excerpt from the dataset is reported in Table~\ref{table:excerpt}. Specifically, for each received IRA packet we log the reception timestamp on the SDR, both in seconds and in milliseconds, the satellite ID, the beam ID, the latitude, longitude, and altitude coordinates of the emitting satellite, and the raw IQ samples included in the IRA packet. As recently discussed by the authors in~\cite{oligeri_wisec_2020}, any IRIDIUM satellite is equipped with a total number of $49$~radios, where $48$ represent the radio of the beams and the remaining one reports the whole satellite ID, characterized by the beam numbered $0$. 
For our work, we further restricted the analysis to \emph{satellite} IRA packets, i.e., the one having beam ID $0$.
\begin{table*}[htbp]
\footnotesize
\centering
\caption{Excerpt of the collected dataset. Latitude and Longitude information anonymized for peer-review.}
\begin{tabular}{ccccccc}
\textbf{Time (s)} & \textbf{Time (ms)} & \textbf{\begin{tabular}[c]{@{}c@{}}Satellite \\ ID\end{tabular}} & \textbf{\begin{tabular}[c]{@{}c@{}}Beam \\ ID\end{tabular}} & \textbf{Latitude} & \textbf{Longitude}& \textbf{\begin{tabular}[c]{@{}c@{}}IQ \\ Samples\end{tabular}} \\ \hline
1580712040 & 000000739 & 115 & 0  & ?            & ?    &    0.03+0.3j, ...    \\
1580712040 & 000004519 & 115 & 0 & ?            & ?      &   0.02-0.4j, ...   \\
1580712040 & 000005059 & 115 & 0 & ?            & ?      &   -0.07+0.8j, ...   \\
1580712040 & 000005599 & 115 & 0 & ?            & ?      &    -0.2-0.4j, ...  \\
1580712040 & 000008839 & 66  & 0  & ?            & ?     &    0.03+0.3j, ...   \\
1580712040 & 000013159 & 66  & 0 & ?            & ?      &    0.03+0.3j, ...  \\
1580712040 & 000013699 & 66  & 0 & ?            & ?      &    0.03+0.3j, ...
\end{tabular}
\label{table:excerpt}
\end{table*}

Finally, we implemented the proposed classification algorithms (\acf{CNN} and autoencoders) in MATLAB R2020a. The training, validation, and testing have been carried out by a server featuring 64 cores, 512GB RAM, and 4 GPUs Nvidia Tesla M40. The collected data will be released open source once the paper will be accepted.

\subsection{Reverse-Engineering IRIDIUM Constellation Parameters}
\label{sec:iridium_parameters}

In this section, we derive important parameters of the IRIDIUM satellite constellation, functional to the subsequent analysis. We consider the \ac{SNR} associated with the collected IQ samples, the waiting time between two consecutive passes of a specific satellite and, finally, the number of IQ samples that can be collected during a single satellite pass.

{\bf Signal-to-Noise Ratio (SNR).} We start the analysis by considering the quality of the collected samples, in terms of \ac{SNR}. Firstly, we compute the received power $P_{rx}$ associated with the IQ samples as in Eq.~\ref{eq:p_rx}:
\begin{equation}
    \label{eq:p_rx}
    %$$
    P_{rx}[dBm] = 10 \cdot \log_{10}(10 \cdot (I^2 + Q^2)),
    %$$ 
\end{equation}
where $I$ and $Q$ are the in-phase and quadrature component of the signal, respectively. Conversely, we evaluated the noise power as in Eq.~\ref{eq:n_dbm}.
\begin{equation}
    \label{eq:n_dbm}
    %$$
    N[dBm] = 10 \cdot \log_{10}(10 \cdot \textrm{var}(I^2 + Q^2)),
    %$$ 
\end{equation}
where $\textrm{var}(\circ)$ is the statistic variance. Finally, the SNR has been computed as in Eq.~\ref{eq:snr}.
\begin{equation}
    \label{eq:snr}
    %$$
    \textrm{SNR}[dB] = P_{rx} - N.
    %$$
\end{equation}

\begin{figure}[htbp]
    \centering
    \includegraphics[width=\columnwidth]{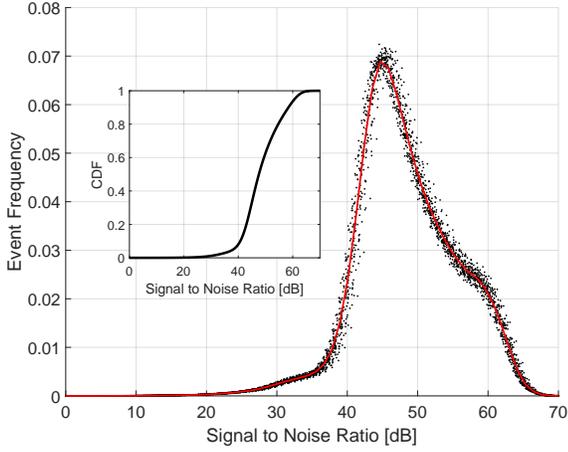}
    \caption{SNR of all the collected measurements. Black dots represent the real SNR values, while the solid red line depicts the best-fit interpolation.}
    \label{fig:rss}
\end{figure}

Black dots in Fig.~\ref{fig:rss} represent the probability density associated with the SNR for all the collected IQ samples, independently of the satellite transmitting source, while the solid red line depicts the best-fit interpolation. We also computed the associated cumulative density function (CDF), as depicted in the inset of Fig.~\ref{fig:rss}. We highlight that the peak is represented by an SNR of about 45dBm, while $90\%$ of the collected samples experience an SNR in the range 40---60 dBm.

% Total number of samples: 102318546
% Samples per sat: 1550281

{\bf Waiting time between consecutive satellite passes.} We also investigate the time an observer (on the ground) has to wait to see again the same satellite. % in the sky more than once. Figure~\ref{fig:waiting_time} shows the probability density function associated with the waiting times, being equal to about $92$, $655$, $633$, $561$, and $535$~minutes, decreasingly sorted by occurring probability.
\begin{figure}[htbp]
    \centering
    \includegraphics[width=\columnwidth]{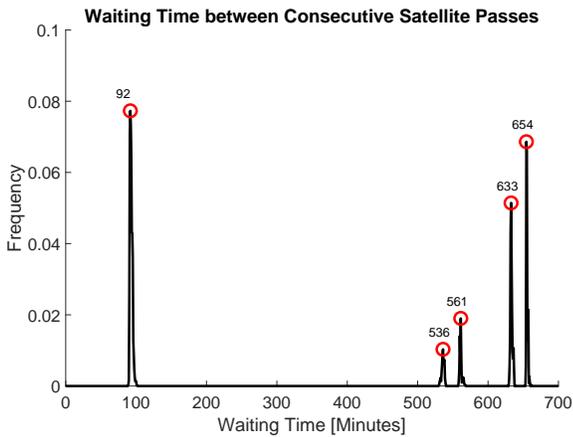}
    \caption{Waiting time among consecutive satellite passes.}
    \label{fig:waiting_time}
\end{figure}

We can explain these results by recalling that a satellite can pass over a specific location in two directions, either north-south or south-north. 
%
%Independently from the direction of the pass, a satellite passes over a single location in the same direction twice, with a period that is %approximately $90$~minutes. After the second consecutive pass in the same direction, the satellite will pass again over the same location in the %opposite direction, after a longer time, that is approximately $560$~minutes. Given that each pass of the satellite is a little displaced compared to %the previous one, there are cases where a satellite passes over a single location only once. In this cases, the waiting time is higher, and it is %approximately $650$~minutes, approx. equal to $560+90$~minutes.
%
%RDP plz check se quanto sotto e' corretto
% Done: Gab
Indeed, each satellite passes over the same location twice every $90$~minutes: up to two consecutive passes can be detected from the same position. Subsequently, after a full Earth revolution, the satellite returns on the same location after about $560$~minutes with opposite direction. Higher waiting times (in Fig.~\ref{fig:waiting_time}), e.g., $560+90 \approx 650$~minutes, are due to passes that have not been detected by the receiver.

%RDP questa ultima misura va spiegata; messa cosi sembrerebbe che ha completato una rotazione NE e poi fatto  dietro-front
%gab: done

\begin{figure}[htbp]
    \centering
    \includegraphics[width=\columnwidth]{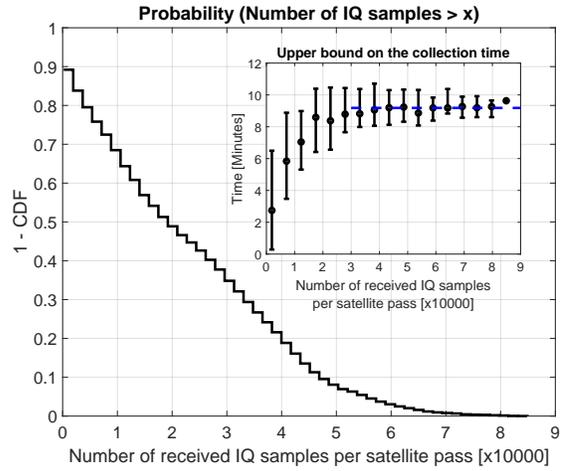}
    \caption{Probability to experience at least $x$ IQ samples in a single satellite pass.}
    \label{fig:iq_sample_per_pass}
\end{figure}

{\bf IQ samples per satellite pass.} Another important parameter for the subsequent analysis is the number of collected IQ samples per satellite pass, i.e., the number of IQ samples that can be collected by a receiver during a single satellite pass. Firstly, we consider the inverse cumulative distribution function associated with the number of received IQ samples ($N$) per satellite pass, as depicted in Fig.~\ref{fig:iq_sample_per_pass}, i.e., $P(N > x)$, where $x$ represents a predefined value of IQ samples. The overall trend is linear up to $50,000$~samples: it is worth noting a probability of $0.7$ and $0.5$ to have at least $10,000$ and $20,000$ samples per satellite pass. 
% Number of captured passes: 83, 92, 94
% Median value: 9.18

The inset of Fig.~\ref{fig:iq_sample_per_pass} shows the time required to collect the IQ samples. For instance, $10,000$ and $20,000$ IQ samples can be collected by satellite passes lasting for $7$ and $8$ minutes, respectively. The satellite passes last for a maximum time of $9$ minutes (median value of the maxima); during this period, we were able to collect between $30,000$ and $80,000$ IQ samples. We explain this wide range of values due to the varying noise conditions during the measurement campaign. Finally, it is worth noting the trend between $0$ and $30,000$ IQ samples, characterized by satellite pass length between $3$ and $8$ minutes. We consider these events to be associated to passes close to the horizon, where the satellite appears %and disappears 
just for a short amount of time.

\subsection{Transmitting-source Authentication via IQ samples}
\label{sec:real_iq_samples}

\begin{figure}[htbp]
    \centering
    \includegraphics[width=\columnwidth]{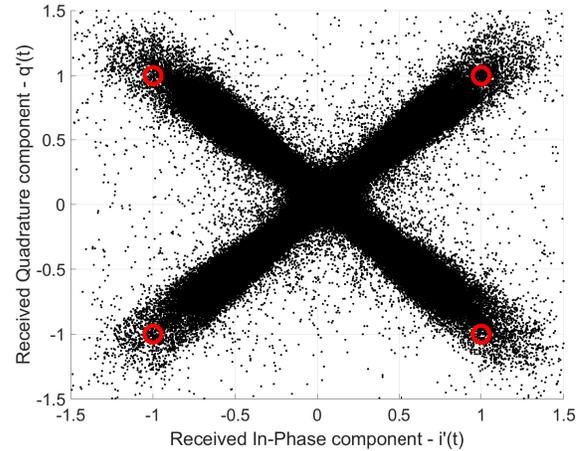}
    \caption{Received In-Phase $i'(t)$ and Quadrature $q'(t)$ components of 679,740 samples from Satellite with ID 7.}
    \label{fig:iq_example}
\end{figure}

Figure~\ref{fig:iq_example} shows the received In-Phase $i'(t)$ and Quadrature $q'(t)$ components of $679,740$ samples gathered from the Satellite with ID $7$. It is worth noting that the ideal IQ constellation (recall Fig.~\ref{fig:bits_to_symbols}) is significantly different from the one experienced in real down-link satellite communications. Red circles in Fig.~\ref{fig:iq_example} highlight the ideal positions of the IQ samples and identify the four Cartesian quadrants adopted for the decision (recall Fig.~\ref{fig:bits_to_symbols}), i.e., received IQ sample (black dot) is mapped to the corresponding red circle as a function of the Cartesian quadrant on which it lies. The received IQ samples are affected by different phenomena that displace their original positions. As for the bit error rate, as long as the samples remain in their intended quadrants, the error rate remains zero.
In this contribution, we are not interested in the link error rate; instead, we focus on the phenomena behind the IQ samples' displacement. In general, a received (satellite) signal is affected by the following phenomena:

\begin{itemize}
    \item {\em Fading.} Iridium satellites are \ac{LEO} satellites, hence located at an height of approximately $780$~Km , thus being affected by a significant signal attenuation. Note that Fig.~\ref{fig:iq_example} is the result of a post-processing amplification, where the samples are stretched to fit the Cartesian plane $[-1, 1] \times [-1, 1]$.
    \item {\em Multipath.} Multipath is caused by multiple replicas of the transmitted signal reaching out the receiver through different paths, thus summing up at the receiver, albeit with different phases. Since the phase shift is random, the attenuation can be arbitrarily large, causing a destructive interference that can significantly affect the signal decoding.
    \item {\em Doppler shift.} Doppler shift represents the change of frequency (shift) of the received signal as a function of the relative speed between the transmitter and the receiver at the ground. The satellite scenario is particularly challenging, since the Doppler shift is maximum when the satellite is at the receiver's horizon, while becoming minimum at the receiver's zenith.
    \item {\em Hardware impairments.} Although mass produced, any two radio transceivers and their electronic components are not identical. Indeed, such discrete components can be affected by small physical differences at micro and nano scale (e.g. material impurity) that are reflected in variations of capacitance, resistance, and inductance, eventually leading to small (almost undetectable) signal artifacts and IQ unbalances. While the cited imperfections do not affect communication performance, they make the transmitted signal unique, thus (theoretically) enabling the identification of the transmitting source. Unfortunately, this is not an easy task, since such small IQ unbalances are hidden by all the previously-discussed phenomena---each of them having a sensitive impact in the IQ unbalancing. In the following, we will discuss an AI-based methodology to detect and extract such imperfections, and we will prove our approach being robust to noise, and able to identify a specific satellite transmitter among the $66$ that make up the Iridium constellation---thus enabling the physical authentication of the transmitting source.
\end{itemize}

\subsection{IQ Samples Pre-processing}
\label{sec:iq_samples_preprocessing}

Noise represents a major challenge when the receiver aims at identifying the transmitting source via the IQ unbalances produced by hardware impairments of the the transmitting device. Over the years, several techniques have been developed to address the above issue, and the vast majority of them achieve great performance. Nevertheless, none of the mentioned techniques considered noisy radio links, e.g., like the satellite wireless channel. 
Indeed, recalling Fig.~\ref{fig:iq_example}, it can be observed that IQ samples do not appear just around the ideal points (red circles), but they spread all over the IQ plane. The ``cross''-like shape can be explained by the lack of signal amplitude normalization in the demodulation chain~\cite{saviogithub}. We will prove how the aforementioned issue does not affect our solution, being effective also for small values of the \ac{SNR} (like the ones of a satellite link). %thus being very general and deployable to any communication link.
\begin{figure*}
    \centering
    \subfloat[Histogram of the IQ samples.]{\includegraphics[width=6.6cm]{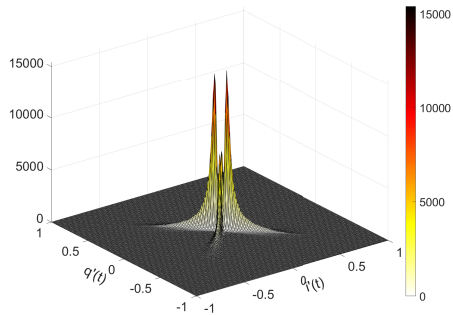}}
    \qquad
    \subfloat[Contour plot (magnified) associated with the histogram of the IQ samples.]{\includegraphics[width=6.6cm]{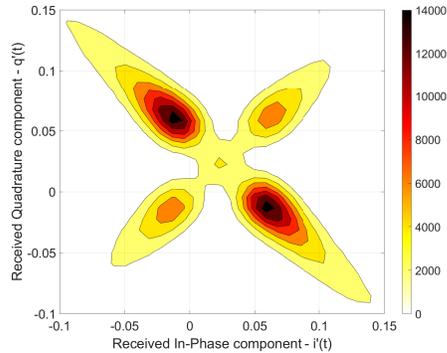}}
    \caption{Image representation of IQ samples.}
    \label{fig:iqtoimage}
\end{figure*}

Our solution involves the adoption of \ac{AI} techniques specifically designed for pattern detection and recognition from images. 
Our approach relies on applying state-of-the-art image pattern recognition techniques to synthetically generated \emph{images of IQ samples}. As previously discussed, hardware impairments generate (consistent, though low intensity) anomalies in the distribution of the IQ samples. Therefore, our intuition is to discriminate between the noise and the anomalies by relying on the more powerful classifiers in the literature.

The aforementioned methodology requires an effective representation of the IQ samples in the image domain. Figure~\ref{fig:iqtoimage} shows how we pre-processed the IQ samples to graphically represent them as images. 
In particular, we sliced the IQ plane into $224 \times 224$ tiles (details on this will be clarified later on), and then we evaluated the deployment of different amounts of IQ samples ($679,740$ from the satellite with ID=7 in Fig.~\ref{fig:iqtoimage}). Subsequently, we computed the bivariate histogram over the aforementioned tiles, i.e., the number of IQ samples belonging to the same tile. Finally, we mapped each value into a grey-scale, i.e., $[0, 255]$, constituting one pixel of our grey image. Therefore, pixels with higher values (white color) represent the tiles with a high number of IQ samples, while pixels with small values (black color) represent tiles with no IQ samples.

A few remarks about Fig.~\ref{fig:iqtoimage}. The figure represents the bipartite histogram associated with the IQ constellation when overlapping multiple IRA messages, each one being constituted by $12$ BPSK symbols (the unique word at the beginning of the frame), $103$ DQPSK symbols (frame content), and $21$ trailing additional DQPSK symbols. Two symbols (second and fourth quadrant) are more likely to appear than the others, due to the modulation overlapping and the trailing sequence (repetition of the same bit values). Finally, Fig.~\ref{fig:iqtoimage}(b) represents the contour plot of the magnification of Fig.~\ref{fig:iqtoimage}(a), where we highlighted the IQ samples density: about $5,500$ samples per tile at the two peaks.

\section{Satellite Authentication Methodologies}
\label{sec:satellite_authentication}

In this section, we describe the proposed methodology to authenticate satellite transmitters.

Specifically, we split the whole IQ samples dataset in three subsets, i.e., \emph{training} ($\mathcal{T}$), \emph{validation} ($\mathcal{V}$), and \emph{testing} ($\mathcal{S}$), each subset accounting for the 60\%, 20\%, and 20\% of the whole dataset, respectively. Moreover, it is worth noting that the number of IQ samples for each satellite is evenly distributed in each subset (i.e., the dataset is balanced by construction). Let us define $\mathcal{D}_s$ the subset of IQ samples from satellite $s$, with $s \in C$ and $C = \{1, \ldots, 66\}$ being the set of satellites in the IRIDIUM constellation. Moreover, let  $\mathcal{D}_s$ be the subset of IQ samples from satellite $s$ and $\mathcal{D}_s = \mathcal{T}_s \cup \mathcal{V}_s \cup \mathcal{S}_s$ where $\mathcal{T}_s$, $\mathcal{V}_s$ and $\mathcal{S}_s$ are the training, validation, and testing subsets associated with the IQ samples from satellite $s$. 

We addressed the physical-layer satellite-authentication problem along two dimensions:
\begin{itemize}
    \item {\bf Multi-class classification.} We aim at being able to correctly authenticate %This approach involves the classification of 
    all the satellites in the constellation. This scenario represents the worst case, involving $66$ equivalent classes. We assume prior knowledge on $\mathcal{T}_s, \forall s \in C$. Moreover, we assume the test subset $\mathcal{S}_x$ to be constituted by IQ samples from the satellite constellation, i.e., $x \in C$---although we do not know to which satellite $s$  the IQ samples belong to.
    \item {\bf Binary classification - One-vs-Rest.} We consider a   candidate satellite $s$, and we combine all the remaining IQ samples (from all the satellites belonging to the constellation), thus obtaining two classes: the class containing the reference satellite $s$, and the one being constituted by all the IQ samples belonging to all the remaining satellites, i.e., $C \setminus \{s\}$. Compared to the previous scenario, this one involves limited prior knowledge, i.e., only $\mathcal{T}_s$, with $s$ being the reference satellite. Moreover, we assume $\mathcal{S}_x$ to be any test subset. Indeed, the algorithm adopted for this categorization returns a \emph{similarity score}, e.g., root mean square, which is used to estimate the similarity of the test subset $\mathcal{S}_x$ against the reference training subset $\mathcal{T}_s$.
\end{itemize}

\begin{table}[]
\caption{Classification strategies.}
\centering
\begin{adjustbox}{width=\columnwidth}
\begin{tabular}{lll}
\multicolumn{1}{l|}{} & \multicolumn{1}{c|}{\textbf{Prior Knowledge}} & \multicolumn{1}{c}{\textbf{Test Subset}} \\ \hline
\multicolumn{1}{l|}{\textbf{Multi-class}}  & \multicolumn{1}{l|}{\begin{tabular}[c]{@{}l@{}}\emph{All satellite} \\ training subsets\end{tabular}}     & \multicolumn{1}{l|}{\begin{tabular}[c]{@{}l@{}}\textit{Any test subset} of satellites \\ belonging to the constellation\end{tabular}} \\ \hline
\multicolumn{1}{l|}{\textbf{One vs Rest}}  & \multicolumn{1}{l|}{\begin{tabular}[c]{@{}l@{}}\emph{Only the reference} \\ training subsets\end{tabular}}     & \multicolumn{1}{l|}{\begin{tabular}[c]{@{}l@{}}\textit{Any test subset} of satellites \\ belonging to the constellation\end{tabular}} \\ \hline
\end{tabular}
\end{adjustbox}
\label{table:ai_algorithm}
\end{table}

Table~\ref{table:ai_algorithm} summarizes our assumptions on the adopted categorization strategies. In the remainder of this paper, we refer to \emph{intra-constellation satellite authentication} as the problem of identifying and  authenticating a satellite by resorting to a multiclass classification tool (see Section~\ref{sec:intra-satellites}). Conversely, we refer to \emph{satellite authentication in the wild} when applying the one-vs-rest classification model (see Section~\ref{sec:satellite_authentication_inthe_wild}).

%% Riferire CNN e autoencoders per i singoli casi

\section{Intra-Constellation Satellite Authentication}
\label{sec:intra-satellites}

In this section, we focus on the intra-constellation satellite authentication scenario. 
Specifically, Section~\ref{sec:convolutional_neural_networks} shows and motivates the deployed CNN, Section~\ref{sec:satellite_authentication_via_cnn} reports details on the application of the described CNN to authenticate IRIDIUM satellite transmitters, while Section~\ref{sec:authentication_subsets} investigates the \ac{CNN} classification performance on subsets of the satellite constellation.

\subsection{Convolutional Neural Network Setup}
\label{sec:convolutional_neural_networks}

In this paper, the multi-class classification task is supported by a \ac{DCNN} based on a Residual Network with $18$ layers (i.e., \emph{ResNet-18}). The original \emph{ResNet-18} has its last fully connected layer composed of $1,000$ neurons (followed by a \textit{softmax} activation function), since it was pretrained on \emph{ImageNet}, a $1,000$-class dataset. Given that our task is to classify $66$~satellites, we replaced the last fully connected \emph{softmax} layer with a fully connected layer composed of 66 neurons only, the number of classes of our dataset. Then, we transferred the set of parameters of the \emph{ResNet-18} convolutional layers to the convolutional layers of our \ac{DCNN}.
As mentioned above, although there were many architectures available in the literature, \emph{ResNet} proved to be the most performing \ac{CNN} by construction, since its structure allows to achieve a higher number of layers, while keeping low the number of parameters~\cite{he2016_cvpr}. 

\begin{figure*}[htbp]
    \centering
    \includegraphics[width=\textwidth]{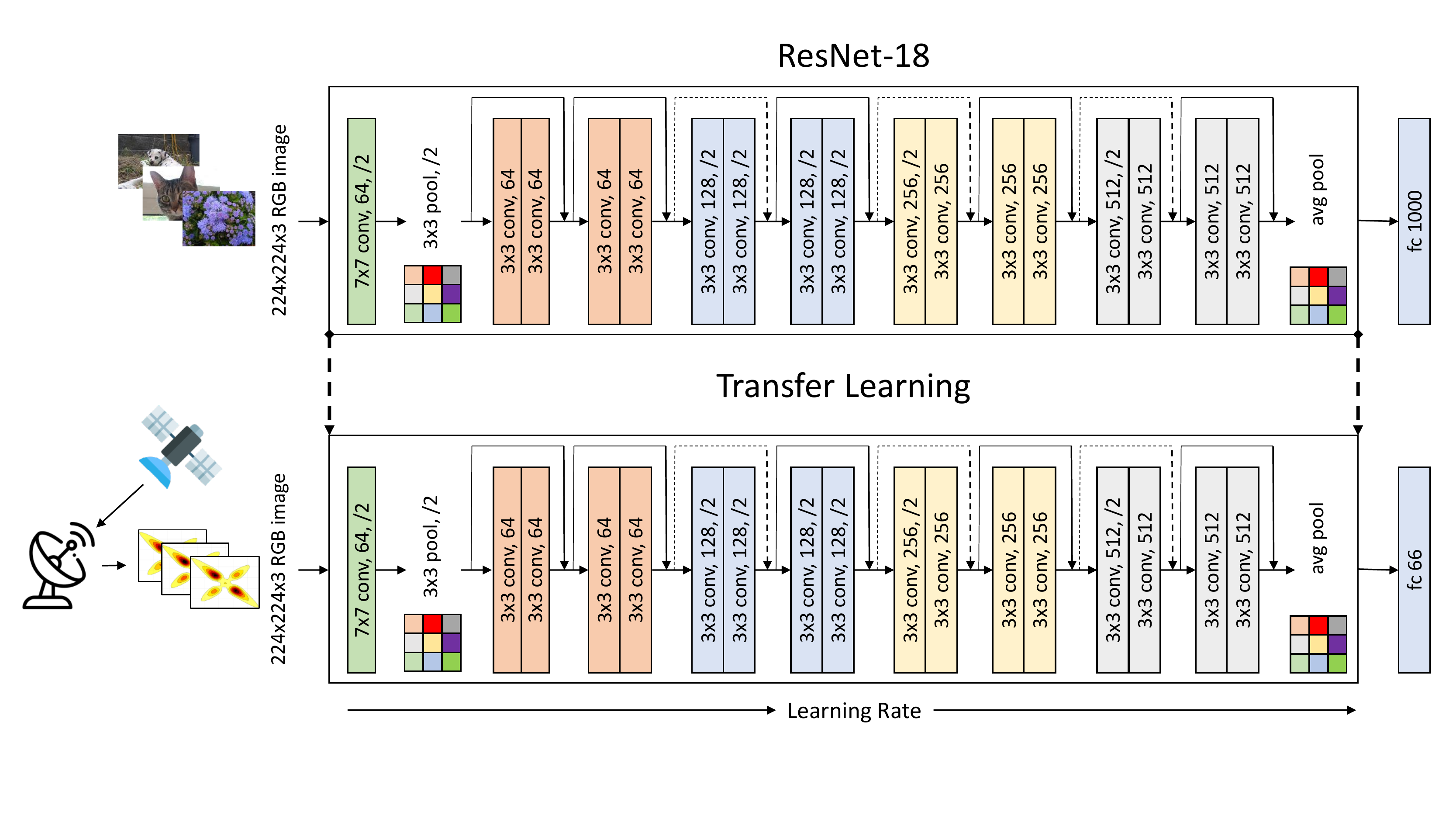}
    \caption{Overview of the proposed architecture. \emph{ResNet-18} pre-trained layers are transferred to our \ac{DCNN}, with the replacement of the fully connected layer (i.e., from $1,000$ neurons to $66$), and the fine-tuning with monotonically increasing learning rate.}
    \label{fig:resnet-18}
\end{figure*}

There are mainly two ways to perform transfer learning in deep neural networks: (i) the fine-tuning approach; and, (ii) the freezing layers approach~\cite{yosinski2014how}. 
The fine-tuning approach requires to retrain (i.e., unfreeze) the whole network parameters, with the classification errors coming from the new training backpropagating to the whole network. 
The freezing layer approach, instead, leaves unchanged (i.e., frozen) most of the transferred feature layers. Generally speaking, when the dataset is small compared to the original one (i.e., the dataset on which the network was pre-trained), the freezing layers approach is suggested, otherwise the fine-tuning approach is the most suitable. However, Yosinki et al. in~\cite{yosinski2014how} showed that the freezing layers approach may lead to a drop in  performance, while the co-adaptation of the features re-learned with the fine-tuning approach prevents this effect. Since it has been observed that the lower layers of a \ac{CNN} are able to detect features that are usually general for each image recognition task (e.g., curves and edges), and that fine-tuning allows to prevent accuracy drops, in this study we rely on a combination of the two approaches. 
Indeed, instead of retraining the network from scratch (i.e., fine-tuning approach) or keeping the layers frozen (i.e., freezing layers approach), we fine-tune the layers of the network with a monotonically increasing learning rate: the deeper the layer in the \ac{CNN}, the higher the learning rate. 
In this way, the parameters of the first layers can still detect common features in images, and we opportunely tune the parameters of the deeper layers in a way to guarantee high accuracy. Figure~\ref{fig:resnet-18} summarizes the proposed architecture.

\subsection{Satellite Authentication via CNN}
\label{sec:satellite_authentication_via_cnn}

\begin{figure}[htbp]
    \centering
    \includegraphics[width=\columnwidth]{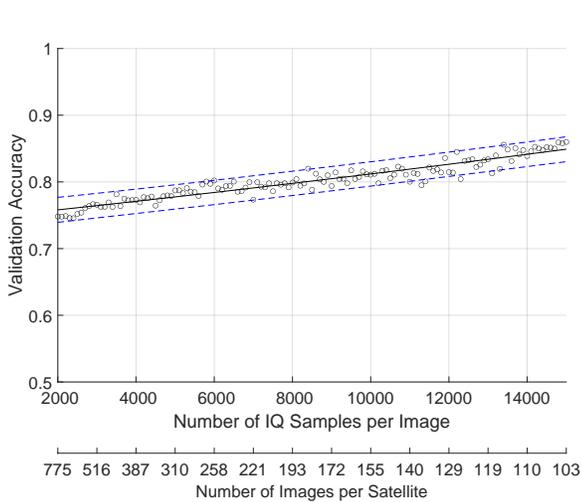}
    \caption{Validation accuracy as a function of the number of IQ samples per image (or number of images per satellite).}
    \label{fig:valacc_samplesimag}
\end{figure}

In this section, we address the problem of authenticating a satellite by classifying the received IQ samples. As discussed in Section~\ref{sec:iq_samples_preprocessing}, IQ samples are pre-processed and converted to $224 \times 224$ greyscale images. Grouping the IQ samples into images involves the following trade-off: on the one hand increasing the number of IQ samples enriches the information possibly conveyed by a single image; on the other hand, the number of available images is reduced smaller, this latter one being the actual input for the classification algorithm that typically performs better as  the size of its input increases.\\* Figure~\ref{fig:valacc_samplesimag} shows the validation accuracy as a function of the number of IQ samples per image (or the number of images per satellite). Each circle in the figure represents the result of a single training and validation process while varying the number of IQ samples per image. Moreover, we recall that for each satellite IQ samples subset % has been previously divided: 
60\% of them have been used for training and 20\% for validation.

The number of IQ samples per image is an important parameter that should be compared with Fig.~\ref{fig:iq_sample_per_pass}. Indeed, the number of IQ samples per image should be matched to a single satellite pass. We could consider waiting for multiple satellites passes, but this approach would involve long waiting times, i.e., at least 92 minutes for the satellite to appear again (recall Fig.~\ref{fig:waiting_time}. Therefore, as a reference parameter, we decided to consider $10,000$ IQ samples per image (leading to 155 images per satellite), guaranteeing a validation accuracy of about $0.83$. Note that the probability to experience at least $10,000$ IQ samples is about $0.7$.

{\bf Testing.} We run $30$ iterations of the training, validation, and testing sequence by randomly choosing the images from the dataset. We computed the mean of the resulting confusion matrices from the testing procedure---results in Appendix. 
The confusion matrix is sorted according to the values in the diagonal, i.e., best performance ($31$) in the top left part of the matrix), being $31$ images ($20\%$ of total 155 images per satellite) the size of the test set for each satellites' image.
\begin{figure}[htbp]
    \centering
    \includegraphics[width=\columnwidth]{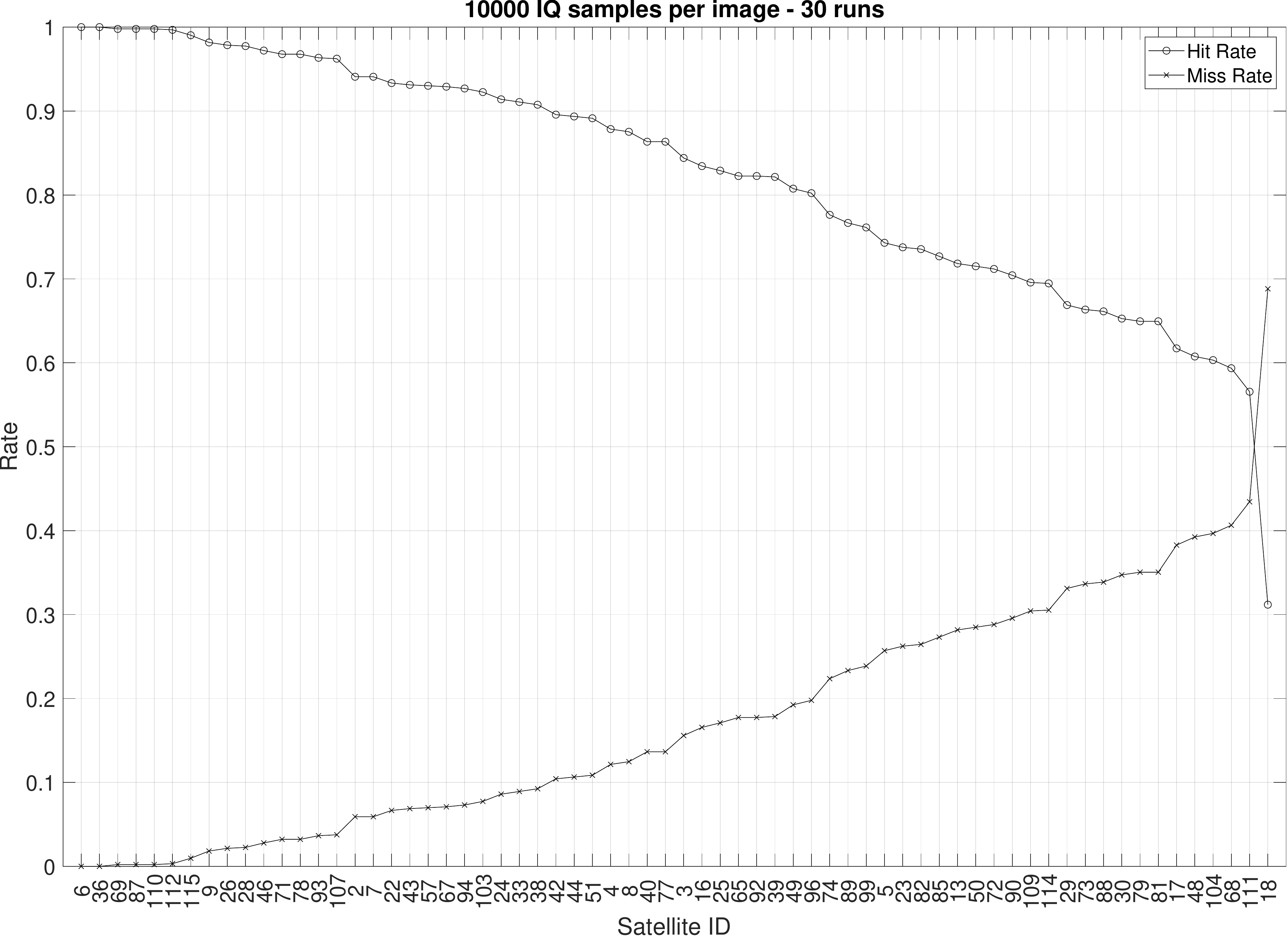}
    \caption{Hit and Miss rates (mean values) for 30 runs of the \ac{CNN} classification algorithm. For each run, we consider the whole training, validation, and testing procedures.}
    \label{fig:cm_10000_acc_err}
\end{figure}

Let us define as the \emph{hit rate} the ratio between the total number of hits (true positive) and the total number of instances (test subset cardinality), yielding:
$$ \textrm{hit\ rate} = \frac{TP}{TP+FN}$$
Moreover, let us define as \emph{miss rate} the ratio between the total number of misses (false negative) and the total number of instances (test subset cardinality), yielding:
$$ \textrm{miss\ rate} = \frac{FN}{TP+FN}$$

Figure~\ref{fig:cm_10000_acc_err} shows the hit and miss rates for each satellite in the IRIDIUM constellation, extracted from the data associated with the aforementioned testing procedure (recall the confusion matrix in Appendix).
%RDP3 la ref all'appendice e' dangling
% Done: Simone

We observe that $24$ satellites (more than $36$\% of the constellation) experience a hit rate higher than $0.9$, while only $4$ satellites have a hit rate less than 0.5. 

\subsection{Authentication of satellite subsets}
\label{sec:authentication_subsets}

Driven by the results of Section~\ref{sec:intra-satellites}, we investigate the \ac{CNN} classification performance on subsets of the satellite constellation. The intuition relies on removing satellites characterized by high miss rates, which are intrinsically difficult to classify, thus constituting a source of mis-classification for the remaining ones. Therefore, we systematically removed the worst satellites (in terms of hit rate) from the dataset, and we subsequently re-evaluated the performance of the classifier.
\begin{figure}[htbp]
    \centering
    \includegraphics[width=\columnwidth]{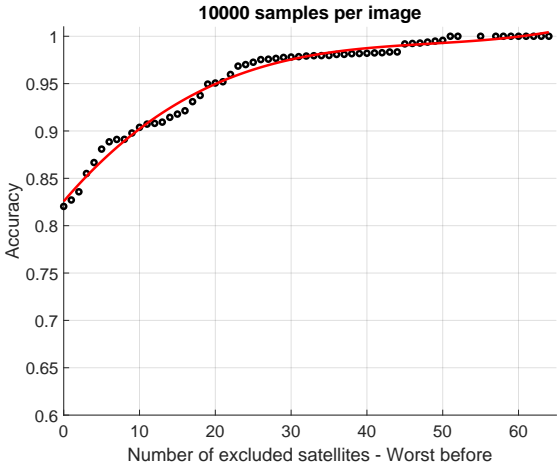}
    \caption{Testing accuracy as a function of the number of excluded satellites. The removed satellites are the ones with worst performance in terms of hit rate. }
    \label{fig:accuracy_exsatellite}
\end{figure}

Figure~\ref{fig:accuracy_exsatellite} shows the accuracy associated with the testing procedure as a function of the number of excluded satellites (the next satellite to be removed is the one with the poorest hit rate among the ones left). 
The analysis confirms that image-based classification of IQ samples is an effective solution. 
Indeed, \ac{CNN} classification guarantees a baseline accuracy above  $0.82$, which can be made arbitrarily high by removing a few satellites---for instance, removing the worst 9 satellites, the accuracy is higher than $0.9$.

\section{Satellite Authentication in the Wild}
\label{sec:satellite_authentication_inthe_wild}

In this section, we undertake the challenge of  authenticating a satellite with minimal prior knowledge, i.e., only a training subset from the satellite to be authenticated. 
Our intuition is to train a model with a reference training subset, and subsequently, to challenge it with a random test subset. Subsequently, we define a metric, i.e., \emph{reproduction error}, and we estimate the deviation of a synthetically-generated subset from the original one. The reproduction error implies a threshold, under which all the samples are considered as belonging to the satellite to be authenticated.

The most suitable class of algorithms for implementing the aforementioned strategy are  \emph{autoencoders}. Indeed, after the training phase, the autoencoders will be biased towards the training subset. Therefore, we expect that a synthetically-generated test subset will be characterized by a higher reproduction error, thus being discarded as not belonging to the satellite to be authenticated. We selected the reproduction error as coincident with the \ac{m.s.e.}. 

In the remainder of this section, we first discuss the architecture of the deployed autoencoders (Section~\ref{sec:autoencoders_auth}). Then, we consider two scenarios: One-vs-Rest (Section~\ref{sec:one-vs_rest}) and One-vs-One (Section~\ref{sec:one-vs_one}). The former undertakes the challenge of authenticating the IQ samples from a reference satellite when compared with IQ samples coming from a set of sources (the other satellites from the constellation). The latter refers to the classification of IQ samples coming from two different sources, i.e., the satellite to be authenticated and another (random) one from the constellation.

We stress that our test subset is constituted by IQ samples belonging to the IRIDIUM constellation, only. We consider this assumption the worst-case scenario for our detection algorithms, i.e., the test subset has the same characteristics of the training subset, in terms of technology, scenario, and noise pattern. Moreover, our solution is agnostic to both the content of the messages (bit-string) and the appearance order of the IQ samples, since we collect and classify the IQ samples independently of their mapping to the bit values.

\subsection{Satellite Authentication via Autoencoders}
\label{sec:autoencoders_auth}

\begin{table}[htbp]
    \caption{Training options of our autoencoder.}
    \label{table:training_options}
    \centering
    \begin{tabular}{|c|c|}
    \hline
    \textbf{Parameter} & \textbf{Value} \\ \hline
    \emph{HiddenSize} & 1,024 \\ \hline
    \emph{MaxEpochs} & 100 \\ \hline
    \emph{EncoderTransferFunction} & \emph{logsig} \\ \hline
    \emph{DecoderTransferFunction} & \emph{logsig} \\ \hline
    \emph{L2WeightRegularization} & 0.001 \\ \hline
    \emph{SparsityRegularization} & 1 \\ \hline
    \emph{SparsityProportion} & 0.05 \\ \hline
    \emph{LossFunction} & \emph{msesparse} \\ \hline
    \emph{TrainingAlgorithm} & \emph{trainscg} \\ \hline
    \emph{ScaleData} & \emph{true} \\ \hline
    \end{tabular}
\end{table}

In this study, we relied on the MATLAB implementation of the \emph{Sparse Autoencoder} to perform the \emph{one-vs-rest} and \emph{one-vs-one} IRIDIUM satellites classification. 
A sparse autoencoder is an autoencoder whose training involves a penalty (also known as sparsity penalty). Several previous works, such as~\cite{makhzani2013k}, observed that classification tasks may see their performance considerably improved  when the representations are learned in a way that encourages sparsity (e.g., by adding a regularizer to the cost function). In the following, we motivate the choice of the training options of our autoencoder---training options are summarized in table~\ref{table:training_options}. \\
\textbf{HiddenSize.} It represents the number of neurons in the hidden layer of the autoencoder. The higher the number of neurons, the higher the risk of overfitting, while the lower the number of neurons, the higher the risk of underfitting. We empirically set the number of neurons to $1,024$ since, for our problem, it was a satisfactory trade-off between the two cited conflicting dimensions.\\
%underfitting and overfitting. \\
\textbf{MaxEpochs.} It is defined as the maximum number of training epochs or iterations. An epoch is defined as a single pass through the training set for all the training examples. We empirically selected the value $50$, since none of the subsequent epochs brought any benefit to the accuracy of our model. \\
\textbf{EncoderTransferFunction.} It represents the linear transfer function of the encoder, i.e., the activation function of the neurons in the hidden layer. In this study, we empirically chose the standard logistic sigmoid function, whose formula is reported in Eq.~\ref{eq:logistic}:
\begin{equation}
\label{eq:logistic}
f(x) = \frac{1}{1 + e^{-x}}
\end{equation}
%where $x$ represents the input to the function and $e$ the base of natural log, respectively. \\
\textbf{DecoderTransferFunction.} We relied on the same logistic sigmoid function as activation function of the decoders neurons. \\
\textbf{L2WeightRegularization.} Generally speaking, regularization is a technique that discourages a model from becoming too complex, so as to avoid overfitting. 
It works on the assumption that smaller weights generate simpler models, and it requires to add a regularization term on the weights of the cost function, to prevent them from growing uncontrollably. The L2 regularization term is defined according to Eq.~\ref{eq:reg}.
\begin{equation}
\label{eq:reg}
\Omega_{w} = \frac{1}{2}\sum\limits_l^L \sum\limits_j^n \sum\limits_i^k w_{ji}^{(l)},
\end{equation}
where $L$ is the number of hidden layers, $n$ is the number of samples, and $k$ is the number of variables in the input data, respectively. 
This term is added to the loss function of the autoencoder with a multiplicator $\lambda$, that we empirically set to $0.001$. \\
\textbf{SparsityRegularization.} Sparsity regularization methods attempt to leverage the assumption that, to be learned, an output variable can be described by a reduced number of variables in the feature space. The goal of these methods is to select the input variables that best describe the output. In the autoencoder context, the sparsity regularization term is represented by the Kullback-Leibler divergence, reported in Eq.~\ref{eq:div}.

\begin{equation}
\label{eq:div}
\Omega_{s} = \sum\limits_{i=1}^{D^{(1)}}KL(\rho || \hat{\rho_i}) = \sum\limits_{i=1}^{D^{(1)}} \rho log\left(\frac{\rho}{\hat{\rho_i}}\right) + (1-\rho)log\left(\frac{1 - \rho}{1 - \hat{\rho_i}}\right),
\end{equation}
where $\rho$ and $\hat{\rho}$ represent two distributions. The Kullback-Leibler divergence allows to measure the differences of two distributions. %(the output is 0 if they have no differences and becomes larger as they diverge from each other). 
Since this term is inserted within the loss function, minimizing the cost function allows to minimize the term, thus eventually forcing the distributions to be similar. \\
The sparsity regularization parameter (namely, $\beta$) allows to control the impact that the sparsity regularizer $\Omega_{s}$ has in the cost function. 
The higher the parameter, the more impact the regularizer has on the cost function. We empirically set this value to $1$. \\
\textbf{SparsityProportion.} It represents the proportion of training examples a neuron reacts to. The lower the value of this parameter, the more each neuron will be specialized (i.e., by giving high output only for a small number of training examples). Generally speaking, the lower the sparsity proportion, the higher the degree of sparsity is. We empirically set the parameter to $0.05$. \\
\textbf{Loss Function.} We relied on the standard mean squared error performance function, with L2 weight and sparsity regularizers loss function ($msesparse$), defined as in Eq.~\ref{eq:mean}.
\begin{equation}
\label{eq:mean}
E = \frac{1}{N}\sum\limits_{n=1}^N \sum\limits_{k=1}^K {(x_{kn} - \hat{x}_{kn})}^2 + \lambda * \Omega_{w} + \beta * \Omega_{s},
\end{equation}

where the first term in the addition represents the mean squared error, $\lambda$ is the coefficient controlling the impact of the $L_2$ regularization term (i.e., $0.001$ in our case), and $\beta$ is the coefficient controlling the impact of the sparsity regularization term (i.e., $1$ in our case). \\
\textbf{TrainingAlgorithm.} We relied on the scaled conjugate gradient descent~\cite{moller1993scaled} ($trainscg$) learning algorithm to train our autoencoder. The algorithm is based on a class of optimization techniques known as conjugate gradient methods, and proved to be more effective and one order of magnitude faster than the standard backpropagation algorithm. \\
\textbf{ScaleData.} This parameter allows to control the rescaling of the input data. For the training to be effective, the range of the input data has to match the one of the transfer function for the decoder. By setting this value, the autoencoder scales the data whenever there is a need for, to optimize the algorithm learning capabilities. 

%qui
\subsection{One-vs-Rest}
\label{sec:one-vs_rest}

\begin{figure}[htbp]
    \centering
    \includegraphics[width=\columnwidth]{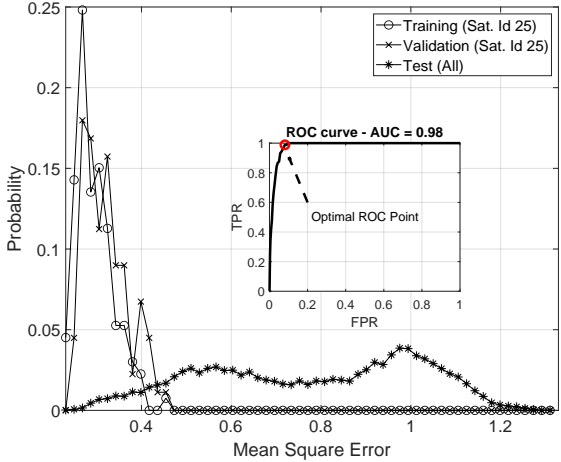}
    \caption{Distribution of the \ac{m.s.e.} for the training, validation, and testing procedures with autoencoders (\emph{One-vs-Rest} scenario for satellite with ID 25). The inset figure highlights the \ac{ROC} curve and the optimal point.}
    \label{fig:mse25}
\end{figure}

In this section, we consider the \emph{One-vs-Rest} scenario: the reference satellite (to be authenticated) versus the rest of the constellation. Figure~\ref{fig:mse25} resumes the results of our methodology for the case of the satellite with $s=25$. 
We trained the autoencoder with the training subset, constituted by the $80$\% of the subset samples from satellite $25$. Then, we used the trained autoencoder to generate a training subset and we estimated the \ac{m.s.e.} between the two subsets, i.e., the original one and the generated one. The circles in Fig.~\ref{fig:mse25} identifies the probability density function associated with the \ac{m.s.e.} computed over the original training subset and the generated one. We performed the same procedure on the validation subset (remaining $20$\% of the samples from satellite $25$), and we computed the probability density function associated with the \ac{m.s.e.} between the original validation subset and the generated one, as depicted by the distribution identified by the crosses in Fig.~\ref{fig:mse25}. It is worth noting that the two distributions (the one associated to the training subset and one associated to the validation subsets) are characterized by the same \ac{m.s.e.}, in the range between $0.2$ and $0.5$.

We applied the same process to a test set. The test set has been constructed by considering all the satellites from the IRIDIUM constellation, but the one with ID $25$. We consider the previous one as the worst-case scenario, since we considered the IQ samples originated from transceivers belonging to the same owner, all of them deployed within a short time delay, % communication technology, 
and hence very likely featuring the same hardware.
Asterisks in Fig.~\ref{fig:mse25} identifies the distribution associated with the \ac{m.s.e.} computed between the generated test and training subset. The test subset is characterized by \ac{m.s.e.} values in the range between $0.7$ and $1.4$, with only a few values less than $0.5$.

By defining a threshold $thr$ in the range between $0.2$ and $1.5$, and assuming as legitimate the \ac{m.s.e.} values less than $thr$, we can experience different \ac{FP} and \ac{FN} events. The trade-off between FP and FN can be evaluated by resorting to the associated \ac{ROC} curve, as shown in in the inset of Fig.~\ref{fig:mse25}, where the \ac{TPR} is evaluated as a function of the \ac{FPR}, with TPR and FPR being $\frac{TP}{TP+FN}$, and $\frac{FP}{FP+TN}$, respectively. In optimal conditions, i.e., $TPR = 1$ and $FPR=0$, the \ac{AUC} should be equal to $1$; in our case, for the developed example related to the satellite with ID $25$, we report an AUC of about $0.98$. 
Finally, we considered the optimal ROC curve, i.e., the best cut-off with the highest TPR and lowest FPR, and we reported this value as the red circle in the inset of Fig.~\ref{fig:mse25}, with coordinates $[0.048, 1]$.

\begin{figure}[htbp]
    \centering
    \includegraphics[width=\columnwidth]{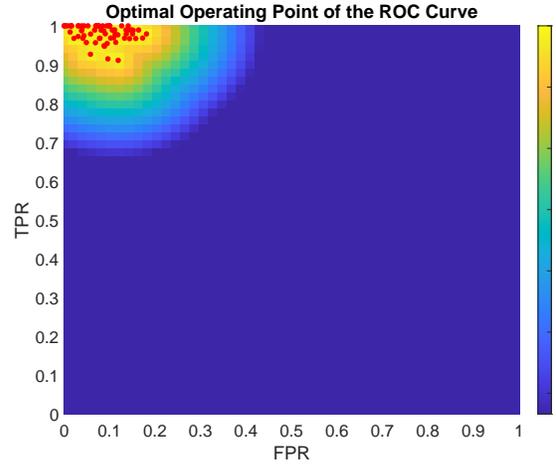}
    \caption{Optimal operating point of the ROC curve for each satellite when testing (with autoencoders) one satellite against the features extracted from the whole constellation dataset (\emph{one-vs-rest}).}
    \label{fig:optimal_point_roc_heatmap}
\end{figure}

We applied the aforementioned procedure for all the satellites in the constellation, thus evaluating the optimal operating point in the ROC curve for each of the investigated satellites. We report the results of our analysis in Fig.~\ref{fig:optimal_point_roc_heatmap}, via a heat-map which reports the minimum distance between each coordinate in the TPR-FPR plane to the optimal points (from the ROC curves). 
The $66$ red dots identifying the optimal operating points of the ROC curves (one per satellite) are very close to each other, and in turn, very close to the optimal point $TPR = 1,~ FPR = 0$.

\begin{figure}[htbp]
    \centering
    \includegraphics[width=\columnwidth]{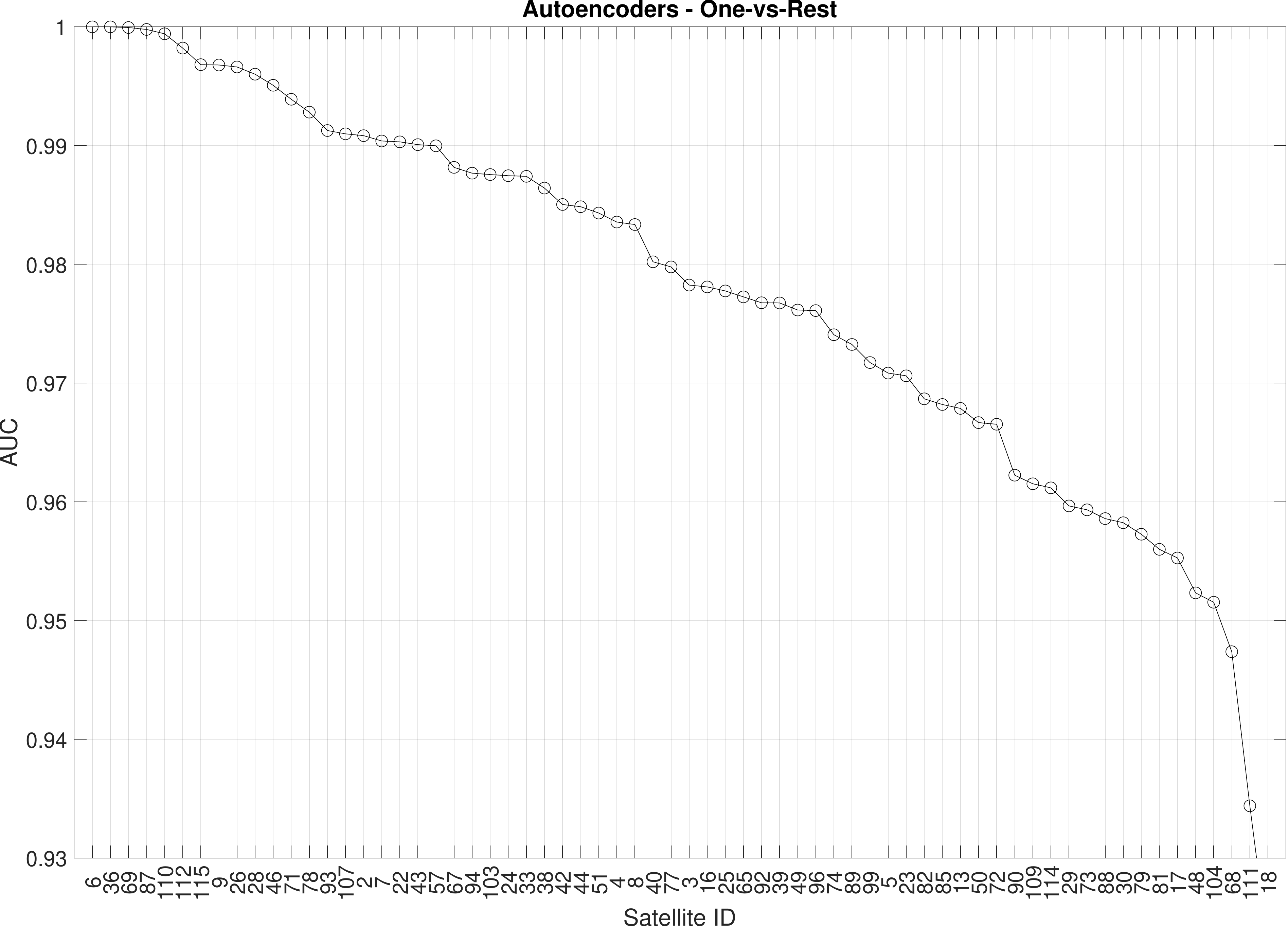}
    \caption{\ac{AUC} for each satellite in the constellation when performing One-vs-Rest classification.}
    \label{fig:auc}
\end{figure}

Finally, we conclude the discussion of the \emph{one-vs-rest} scenario by considering the AUC for each of the satellite in the constellation. Figure~\ref{fig:auc} shows the sorted AUC values for all the satellites in the IRIDIUM constellation. AUC values are characterized by very high values (greater than $0.93$), proving the effectiveness of the proposed solution.

\subsection{One-vs-One}
\label{sec:one-vs_one}
\begin{figure}[htbp]
    \centering
    \includegraphics[width=\columnwidth]{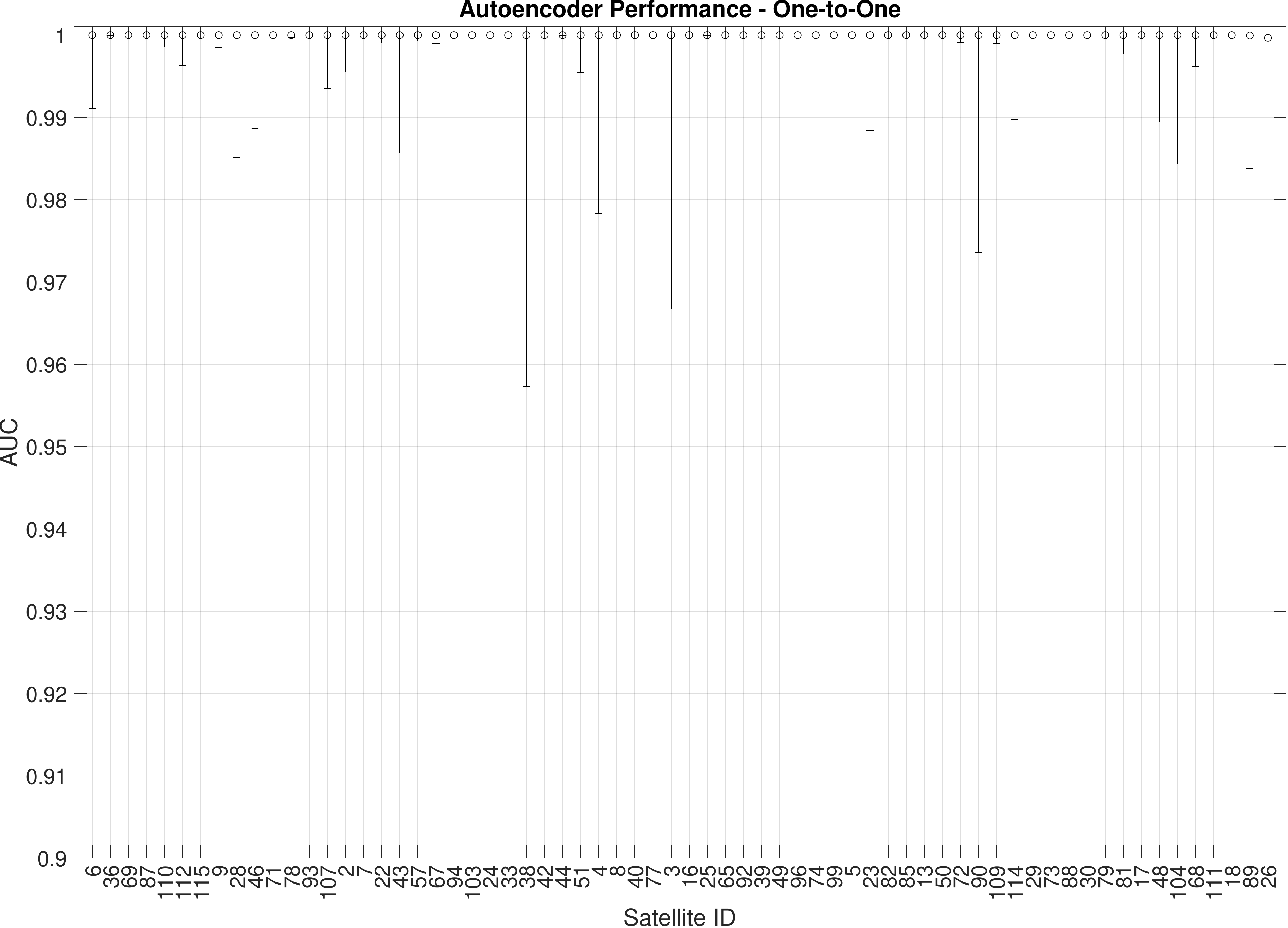}
    \caption{Error-bars (quantile $5$, $50$, and $95$) associated with the \ac{AUC} for each satellite in the constellation, when performing \emph{One-vs-One} classification.}
    \label{fig:autoencoders}
\end{figure}

In this section, we consider the \emph{One-vs-One} scenario: the reference satellite (to be authenticated) versus each one of the satellites in the constellation. We followed the same methodology of Section~\ref{sec:one-vs_rest}, by considering the generation of a training and test subset and their comparison in terms of \ac{m.s.e.} values. 
Finally, we considered different thresholds, and we evaluated the AUC for each satellite pair in the IRIDIUM constellation. Indeed, for each considered reference satellite, we evaluated $66$ classifications and the related AUC. Figure~\ref{fig:autoencoders} shows the error-bars (quantile $95$, $50$, and $5$) associated with each considered reference satellite. We adopted the same order as before, i.e., satellites are sorted by performance (best on the left) considering the median value. We observe that the quantile $95$ and the median are coincident and equal to $1$ for almost all the satellites, while only few satellites are characterized by a quantile $5$ below $0.99$. This is due to a few satellite-to-satellite classifications experiencing lower performance, but still characterized by AUC values greater than $0.96$.

\section{Conclusion}
\label{sec:conclusion}

We presented \proto, a methodology to achieve physical-layer authentication of satellite transmitters by harnessing the power of deep learning classifiers, such as \acp{CNN} and autoencoders, applied to the IQ samples generated by transmitters. We are the first ones, to the best of our knowledge, to prove 
%---for the first time---
that radio fingerprinting can be achieved even in the satellite domain---in particular, for LEO constellations---characterized by high attenuation, multi-path fading, strong Doppler effect, and short link duration.

We investigated the challenges associated with two scenarios: (i) intra-satellite classification; and, (ii) satellite classification in the wild.
We validated our methodology on a   dataset generated from a real measurement campaign, involving more than 100M IQ samples collected  from the IRIDIUM constellation. 
Through a careful adaptation and tuning of the discussed deep learning classifiers, we are able to achieve a classification accuracy that spans between $0.8$ and $1$, depending on the scenario assumptions.

We believe that the novelty of the introduced scenarios, the detailed methodology, the performance achieved by our solution and, finally, the publicly-available dataset, will pave the way for future research in the area.

%\section*{Acknowledgements}
%This publication was made possible by NPRP grants NPRP12S-0125-190013, NPRP-S-11-0109-180242, and NPRP X-063-1-014 from the Qatar National Research Fund (a %member of Qatar Foundation). The findings achieved herein are solely the responsibility of the authors.

\balance
\bibliographystyle{IEEEtran}
\bibliography{iridium_fingerprinting}

\begin{thebibliography}{10}
\providecommand{\url}[1]{#1}
\csname url@samestyle\endcsname
\providecommand{\newblock}{\relax}
\providecommand{\bibinfo}[2]{#2}
\providecommand{\BIBentrySTDinterwordspacing}{\spaceskip=0pt\relax}
\providecommand{\BIBentryALTinterwordstretchfactor}{4}
\providecommand{\BIBentryALTinterwordspacing}{\spaceskip=\fontdimen2\font plus
\BIBentryALTinterwordstretchfactor\fontdimen3\font minus
  \fontdimen4\font\relax}
\providecommand{\BIBforeignlanguage}[2]{{%
\expandafter\ifx\csname l@#1\endcsname\relax
\typeout{** WARNING: IEEEtran.bst: No hyphenation pattern has been}%
\typeout{** loaded for the language `#1'. Using the pattern for}%
\typeout{** the default language instead.}%
\else
\language=\csname l@#1\endcsname
\fi
#2}}
\providecommand{\BIBdecl}{\relax}
\BIBdecl

\bibitem{wang2016_commag}
X.~{Wang}, P.~{Hao}, and L.~{Hanzo}, ``{Physical-layer authentication for
  wireless security enhancement: current challenges and future developments},''
  \emph{IEEE Communications Magazine}, vol.~54, no.~6, pp. 152--158, 2016.

\bibitem{xiao2007_icc}
L.~{Xiao}, L.~{Greenstein}, N.~{Mandayam}, and W.~{Trappe}, ``{Fingerprints in
  the Ether: Using the Physical Layer for Wireless Authentication},'' in
  \emph{2007 IEEE International Conference on Communications}, 2007, pp.
  4646--4651.

\bibitem{xu2016_comst}
Q.~{Xu}, R.~{Zheng}, W.~{Saad}, and Z.~{Han}, ``{Device Fingerprinting in
  Wireless Networks: Challenges and Opportunities},'' \emph{IEEE Communications
  Surveys \& Tutorials}, vol.~18, no.~1, pp. 94--104, 2016.

\bibitem{ibrahim2020_tecs}
O.~A. Ibrahim, S.~Sciancalepore, G.~Oligeri, and R.~{Di Pietro}, ``{MAGNETO:
  Fingerprinting USB Flash Drives via Unintentional Magnetic Emissions},''
  2020.

\bibitem{schmidt2016_csur}
D.~Schmidt, K.~Radke, S.~Camtepe, E.~Foo, and M.~Ren, ``{A survey and analysis
  of the GNSS spoofing threat and countermeasures},'' \emph{ACM Computing
  Surveys (CSUR)}, vol.~48, no.~4, pp. 1--31, 2016.

\bibitem{soltanieh2020_jrfi}
N.~{Soltanieh}, Y.~{Norouzi}, Y.~{Yang}, and N.~C. {Karmakar}, ``{A Review of
  Radio Frequency Fingerprinting Techniques},'' \emph{IEEE Journal of Radio
  Frequency Identification}, vol.~4, no.~3, pp. 222--233, 2020.

\bibitem{oligeri_wisec_2019}
G.~Oligeri, S.~Sciancalepore, O.~A. Ibrahim, and R.~Di~Pietro, ``{Drive Me Not:
  GPS Spoofing Detection via Cellular Network: (Architectures, Models, and
  Experiments)},'' in \emph{Proceedings of the 12th Conference on Security and
  Privacy in Wireless and Mobile Networks}, ser. WiSec '19.\hskip 1em plus
  0.5em minus 0.4em\relax Association for Computing Machinery, 2019, p.
  12–22.

\bibitem{oligeri_wisec_2020}
G.~Oligeri, S.~Sciancalepore, and R.~Di~Pietro, ``{GNSS spoofing detection via
  opportunistic IRIDIUM signals},'' in \emph{Proceedings of the 13th ACM
  Conference on Security and Privacy in Wireless and Mobile Networks}, 2020,
  pp. 42--52.

\bibitem{zhou2019_cns}
X.~{Zhou}, A.~{Hu}, G.~{Li}, L.~{Peng}, Y.~{Xing}, and J.~{Yu}, ``{Design of a
  Robust RF Fingerprint Generation and Classification Scheme for Practical
  Device Identification},'' in \emph{IEEE Conference on Communications and
  Network Security (CNS)}, 2019, pp. 196--204.

\bibitem{yu2019_wimob}
J.~{Yu}, A.~{Hu}, F.~{Zhou}, Y.~{Xing}, Y.~{Yu}, G.~{Li}, and L.~{Peng},
  ``{Radio Frequency Fingerprint Identification Based on Denoising
  Autoencoders},'' in \emph{International Conference on Wireless and Mobile
  Computing, Networking and Communications (WiMob)}, 2019, pp. 1--6.

\bibitem{sankhe2020_tccn}
K.~{Sankhe}, M.~{Belgiovine}, F.~{Zhou}, L.~{Angioloni}, F.~{Restuccia},
  S.~{D’Oro}, T.~{Melodia}, S.~{Ioannidis}, and K.~{Chowdhury}, ``{No Radio
  Left Behind: Radio Fingerprinting Through Deep Learning of Physical-Layer
  Hardware Impairments},'' \emph{IEEE Transactions on Cognitive Communications
  and Networking}, vol.~6, no.~1, pp. 165--178, 2020.

\bibitem{ying2019_cns}
X.~{Ying}, J.~{Mazer}, G.~{Bernieri}, M.~{Conti}, L.~{Bushnell}, and
  R.~{Poovendran}, ``{Detecting ADS-B Spoofing Attacks Using Deep Neural
  Networks},'' in \emph{IEEE Conference on Communications and Network Security
  (CNS)}, 2019, pp. 187--195.

\bibitem{shawabka2020_infocom}
A.~{Al-Shawabka}, F.~{Restuccia}, S.~{D’Oro}, T.~{Jian}, B.~{Costa Rendon},
  N.~{Soltani}, J.~{Dy}, S.~{Ioannidis}, K.~{Chowdhury}, and T.~{Melodia},
  ``{Exposing the Fingerprint: Dissecting the Impact of the Wireless Channel on
  Radio Fingerprinting},'' in \emph{IEEE INFOCOM 2020 - IEEE Conference on
  Computer Communications}, 2020, pp. 646--655.

\bibitem{zhuang2018_asiaccs}
Z.~Zhuang, X.~Ji, T.~Zhang, J.~Zhang, W.~Xu, Z.~Li, and Y.~Liu, ``{FBSleuth:
  Fake Base Station Forensics via Radio Frequency Fingerprinting},'' in
  \emph{Proceedings of the 2018 on Asia Conference on Computer and
  Communications Security}, ser. ASIACCS '18, 2018, p. 261–272.

\bibitem{wang2020_infocom}
S.~Wang, L.~Peng, H.~Fu, A.~Hu, and X.~Zhou, ``{A Convolutional Neural
  Network-Based RF Fingerprinting Identification Scheme for Mobile Phones},''
  in \emph{IEEE INFOCOM 2020-IEEE Conference on Computer Communications
  Workshops (INFOCOM WKSHPS)}.\hskip 1em plus 0.5em minus 0.4em\relax IEEE,
  2020, pp. 115--120.

\bibitem{sankhe2019_infocom}
K.~{Sankhe}, M.~{Belgiovine}, F.~{Zhou}, S.~{Riyaz}, S.~{Ioannidis}, and
  K.~{Chowdhury}, ``{ORACLE: Optimized Radio clAssification through
  Convolutional neuraL nEtworks},'' in \emph{IEEE INFOCOM 2019 - IEEE
  Conference on Computer Communications}, 2019, pp. 370--378.

\bibitem{jian2020_iotmag}
T.~Jian, B.~C. Rendon, E.~Ojuba, N.~Soltani, Z.~Wang, K.~Sankhe, A.~Gritsenko,
  J.~Dy, K.~Chowdhury, and S.~Ioannidis, ``{Deep Learning for RF
  Fingerprinting: A Massive Experimental Study},'' \emph{IEEE Internet of
  Things Magazine}, vol.~3, no.~1, pp. 50--57, 2020.

\bibitem{jafari2018_milcom}
H.~{Jafari}, O.~{Omotere}, D.~{Adesina}, H.~{Wu}, and L.~{Qian}, ``{IoT Devices
  Fingerprinting Using Deep Learning},'' in \emph{IEEE Military Communications
  Conference (MILCOM)}, 2018, pp. 1--9.

\bibitem{bassey2019_fmec}
J.~{Bassey}, D.~{Adesina}, X.~{Li}, L.~{Qian}, A.~{Aved}, and T.~{Kroecker},
  ``{Intrusion Detection for IoT Devices based on RF Fingerprinting using Deep
  Learning},'' in \emph{International Conference on Fog and Mobile Edge
  Computing (FMEC)}, 2019, pp. 98--104.

\bibitem{balakrishnan2020_tifs}
S.~{Balakrishnan}, S.~{Gupta}, A.~{Bhuyan}, P.~{Wang}, D.~{Koutsonikolas}, and
  Z.~{Sun}, ``{Physical Layer Identification Based on Spatial–Temporal Beam
  Features for Millimeter-Wave Wireless Networks},'' \emph{IEEE Transactions on
  Information Forensics and Security}, vol.~15, pp. 1831--1845, 2020.

\bibitem{foruhandeh2020_wisec}
M.~Foruhandeh, A.~Z. Mohammed, G.~Kildow, P.~Berges, and R.~Gerdes, ``{Spotr:
  GPS Spoofing Detection via Device Fingerprinting},'' in \emph{Proceedings of
  the 13th ACM Conference on Security and Privacy in Wireless and Mobile
  Networks}, ser. WiSec '20, 2020, p. 242–253.

\bibitem{rappaport}
T.~Rappaport, \emph{Wireless Communications: Principles and Practice},
  2nd~ed.\hskip 1em plus 0.5em minus 0.4em\relax USA: Prentice Hall PTR, 2001.

\bibitem{goodfellow_deep_learning}
I.~Goodfellow, Y.~Bengio, and A.~Courville, \emph{Deep Learning}.\hskip 1em
  plus 0.5em minus 0.4em\relax MIT Press, 2016,
  \url{http://www.deeplearningbook.org}.

\bibitem{kingma2019introduction}
D.~P. Kingma and M.~Welling, ``An introduction to variational autoencoders,''
  \emph{arXiv preprint arXiv:1906.02691}, 2019.

\bibitem{ieracitano2020neurocomputing}
\BIBentryALTinterwordspacing
C.~Ieracitano, A.~Adeel, F.~C. Morabito, and A.~Hussain, ``A novel statistical
  analysis and autoencoder driven intelligent intrusion detection approach,''
  \emph{Neurocomputing}, vol. 387, pp. 51 -- 62, 2020. [Online]. Available:
  \url{http://www.sciencedirect.com/science/article/pii/S0925231219315759}
\BIBentrySTDinterwordspacing

\bibitem{nazir2020autoencoder}
S.~Nazir, S.~Patel, and D.~Patel, ``Autoencoder based anomaly detection for
  scada networks,'' \emph{International Journal of Artificial Intelligence and
  Machine Learning (IJAIML)}, 2020.

\bibitem{yang2020ieeeifip}
K.~{Yang}, J.~{Zhang}, Y.~{Xu}, and J.~{Chao}, ``Ddos attacks detection with
  autoencoder,'' in \emph{NOMS 2020 - 2020 IEEE/IFIP Network Operations and
  Management Symposium}, 2020, pp. 1--9.

\bibitem{raponi2020sound}
S.~Raponi, I.~Ali, and G.~Oligeri, ``{Sound of Guns: Digital Forensics of Gun
  Audio Samples meets Artificial Intelligence},'' \emph{arXiv preprint
  arXiv:2004.07948}, 2020.

\bibitem{liu2017survey}
W.~Liu, Z.~Wang, X.~Liu, N.~Zeng, Y.~Liu, and F.~E. Alsaadi, ``A survey of deep
  neural network architectures and their applications,'' \emph{Neurocomputing},
  vol. 234, pp. 11--26, 2017.

\bibitem{he2016_cvpr}
K.~{He}, X.~{Zhang}, S.~{Ren}, and J.~{Sun}, ``{Deep Residual Learning for
  Image Recognition},'' in \emph{IEEE Conference on Computer Vision and Pattern
  Recognition (CVPR)}, 2016, pp. 770--778.

\bibitem{pratt1999_comst}
S.~R. Pratt, R.~A. Raines, C.~E. Fossa, and M.~A. Temple, ``{An operational and
  performance overview of the IRIDIUM low earth orbit satellite system},''
  \emph{IEEE Communications Surveys}, vol.~2, no.~2, pp. 2--10, 1999.

\bibitem{iridium_iot}
{IRIDIUM Corp.}, ``{Iridium's Internet of Things - Connect to a World of IoT
  Possibilities},'' \url{https://www.iridium.com/solutions/iot/}, February
  2020, accessed: 2020-08-25.

\bibitem{iridium_antenna}
{Beam Communications}, ``{Iridium Beam Active Antenna (RST740)},''
  \url{https://www.beamcommunications.com/satellite/76-iridium-beam-active-antenna},
  2020, (Accessed: 2020-09-07).

\bibitem{ettus}
{Ettus Research}, ``{USRP X310},''
  \url{https://www.ettus.com/all-products/x310-kit/}, 2020, (Accessed:
  2020-09-07).

\bibitem{ubx}
------, ``{UBX160 Daughterboard},''
  \url{https://www.ettus.com/product/details/UBX160}, 2020, (Accessed:
  2020-09-07).

\bibitem{iridiumgr}
C.~C.~C. München, ``Gnuradio iridium out of tree module,''
  \url{https://github.com/muccc/gr-iridium}, Sep. 2019, accessed: 2020-09-07.

\bibitem{iridium-toolkit}
------, ``Simple toolkit to decode iridium signals,''
  \url{https://github.com/muccc/iridium-toolkit}, Sep. 2019, accessed:
  2020-09-07.

\bibitem{saviogithub}
S.~Sciancalepore, ``Weird patterns in i/q values,''
  \url{https://github.com/muccc/gr-iridium/issues/48\#issuecomment-657152591},
  2020, [Online; accessed 26-August-2020].

\bibitem{yosinski2014how}
J.~Yosinski, J.~Clune, Y.~Bengio, and H.~Lipson, ``How transferable are
  features in deep neural networks?'' in \emph{Proceedings of the 27th
  International Conference on Neural Information Processing Systems - Volume
  2}, ser. NIPS'14.\hskip 1em plus 0.5em minus 0.4em\relax Cambridge, MA, USA:
  MIT Press, 2014, p. 3320–3328.

\bibitem{makhzani2013k}
A.~Makhzani and B.~Frey, ``K-sparse autoencoders,'' \emph{arXiv preprint
  arXiv:1312.5663}, 2013.

\bibitem{moller1993scaled}
M.~F. M{\o}ller, ``A scaled conjugate gradient algorithm for fast supervised
  learning,'' \emph{Neural networks}, vol.~6, no.~4, pp. 525--533, 1993.

\end{thebibliography}

%\clearpage

% \begin{appendices}

% \section{Prova}

% %\label{sec:appendix}
% %\includepdf[angle=90,scale=0.8]{figures/cm_10000.pdf}
% \begin{figure*}[htbp]
%     \centering
%     \caption{}
%     %\vspace*{-10mm}
%     \includegraphics[scale=0.15, angle=90]{figures/cm_10000.pdf}
%     \label{TODO}
% \end{figure*}
% \end{appendices}

\newpage
\section*{Appendix}
%\label{sec:appendix}
%\begin{appendices}
\noindent\begin{minipage}{\textwidth}
    \centering
    %\captionof{figure}{Big Figure}
    %\includegraphics[height=0.99\vsize, width=0.99\hsize]{figures/cm_10000.pdf}
    \includegraphics[scale=0.64, angle=90]{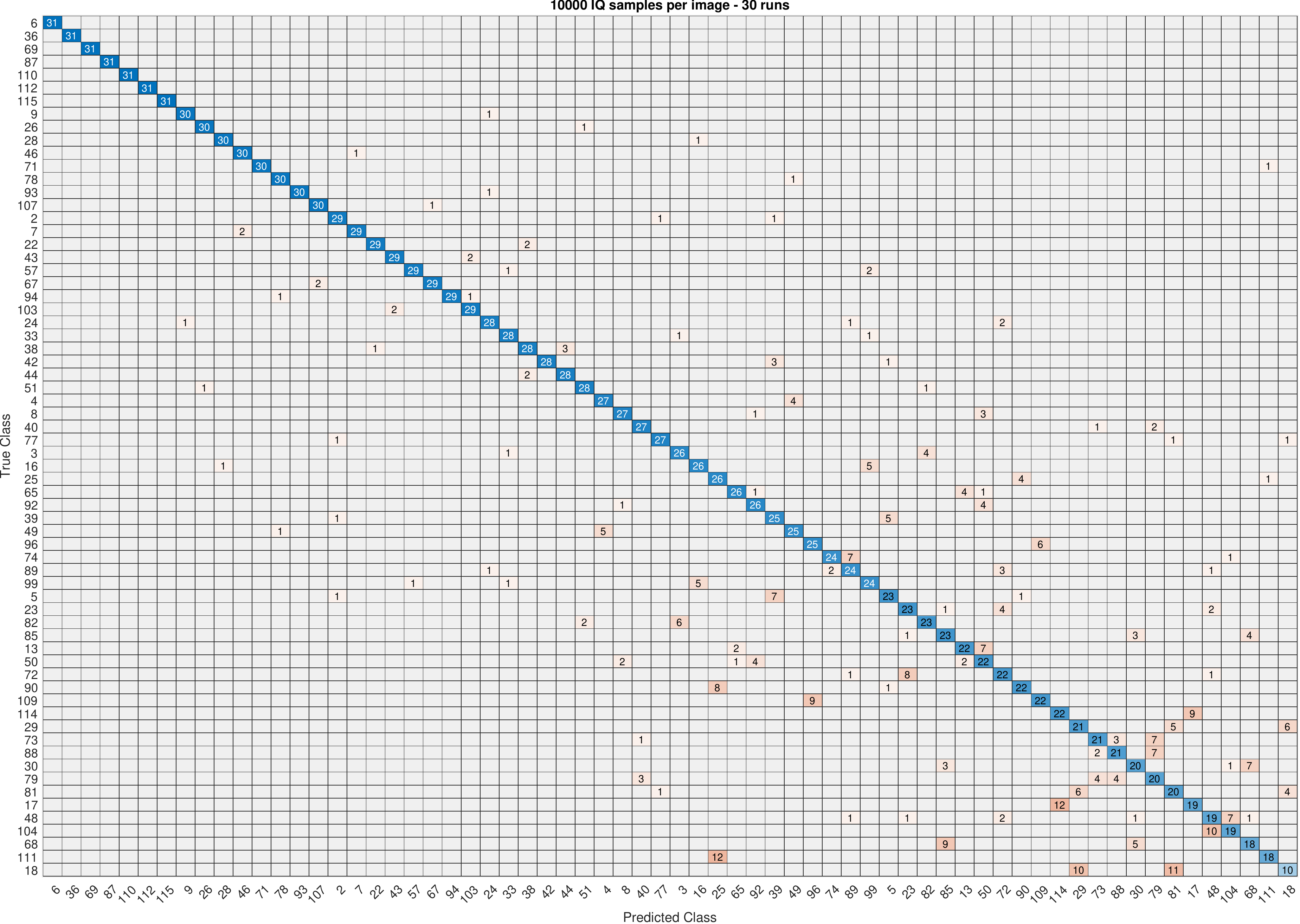}
\end{minipage}
%\end{appendices}

\end{document}